\documentclass[prl,twocolumn,twoside,showpacs,floatfix,superscriptaddress]{revtex4-1}

\usepackage{graphicx,color,rotating}
\usepackage{amsmath,amssymb,bm}
\usepackage{dcolumn}

\usepackage{times,txfonts}

\usepackage{multirow}

\usepackage{pifont}
\usepackage{dsfont}

\usepackage{adjustbox}

\definecolor{FGViolet}{rgb}{0.61,0.32,0.61}
\definecolor{FGDarkBlue}{rgb}{0,0,0.6}
\definecolor{FGBlue}{rgb}{0,0,0.8}
\definecolor{FGLightBlue}{rgb}{0.2, 0.6, 0.8}
\definecolor{FGGreen}{rgb}{0.2,0.7,0.2}
\definecolor{FGLightGreen}{rgb}{0.4,1,0.4}
\definecolor{FGYellow}{rgb}{1,0.95,0}
\definecolor{FGOrange}{rgb}{0.95,0.5,0.1}
\definecolor{FGRed}{rgb}{0.8,0,0}
\definecolor{FGWhite}{rgb}{1,1,1}
\definecolor{FGLightGray}{rgb}{0.8,0.8,0.8}
\definecolor{FGGray}{rgb}{0.5,0.5,0.5}
\definecolor{FGDarkGray}{rgb}{0.3,0.3,0.3}
\definecolor{FGBlack}{rgb}{0,0,0}

\newcommand{\beq}{\begin{equation}}
\newcommand{\eeq}{\end{equation}}
\newcommand{\beqn}{\begin{eqnarray}}
\newcommand{\eeqn}{\end{eqnarray}}

\makeatletter
\renewcommand{\p@subsection}{}
\renewcommand{\p@subsubsection}
\makeatother

\makeatletter
\newcommand{\doublewidetilde}[1]{{%
  \mathpalette\double@widetilde{#1}%
}}
\newcommand{\double@widetilde}[2]{%
  \sbox\z@{$\m@th#1\widetilde{#2}$}%
  \ht\z@=.9\ht\z@
  \widetilde{\box\z@}%
}
\makeatother
\usepackage{titlesec}
\titlespacing\section{0pt}{24pt plus 4pt minus 2pt}{6pt plus 2pt minus 2pt}
\titlespacing\subsection{0pt}{12pt plus 4pt minus 2pt}{0pt plus 2pt minus 2pt}

\begin{document}

\title{Structure and reactions of $^{11}$Be:
many-body basis for single-neutron halo }
\author{F. Barranco}
\affiliation{Departamento de F\`isica Aplicada III,
Escuela Superior de Ingenieros, Universidad de Sevilla, Camino de los Descubrimientos, 	Sevilla, Spain}	
\author{G. Potel}
\affiliation{National Superconducting Cyclotron Laboratory, Michigan State University, East Lansing, Michigan 48824, USA}
\author{R.A. Broglia}	
\affiliation{ The Niels Bohr Institute, University of Copenhagen, 
DK-2100 Copenhagen, Denmark }
\affiliation{Dipartimento di Fisica, Universit\`a degli Studi Milano,
Via Celoria 16, 
I-20133 Milano, Italy }
\author{E. Vigezzi}
\affiliation{INFN Sezione di  Milano,
Via Celoria 16, 
I-20133 Milano, Italy }

\begin{abstract}
The exotic nucleus $^{11}$Be has been extensively studied and much experimental information 
is available on the structure of this system. Treating, within the framework of empirically
renormalised nuclear
field theory $(NFT)_{ren}$ in both configuration and 3D-space, the mixing of bound and continuum single-particle ($sp$) states through the coupling
to collective  particle-hole  (p,h) and pairing vibrations of the $^{10}$Be core, as well as Pauli principle
acting not only between the particles explictely considered and those participating in the collective states,
but also between fermions involved in two-phonon virtual states
 it is possible, for the first time, to simultaneously  
and quantitatively account for
  the energies of the $1/2^+,1/2^-$  low-lying states, the centroid and line shape of the 
 $5/2^+$ resonance,   the one-nucleon stripping and
pickup absolute differential cross sections involving  $^{11}$Be as either target or residual nucleus,
and the dipole transitions connecting the $1/2^+$ and $1/2^-$ parity inverted levels 
as well as the charge radius, thus providing 
a unified and exhaustive characterisation of  
the many-body effects which are at the basis of this 
paradigmatic one-neutron halo system.

\end{abstract}

 \maketitle
\date{today}



At the basis of nuclear structure, one finds the unification (A. Bohr and Mottelson ) of the shell model (Mayer and Jensen)
and the collective model (N. Bohr and Wheeler). One nucleon moving outside closed shell constitutes 
a unique laboratory to test the validity of the above picture in systems which are at the limit of stability of neutron 
or proton excess, drip lines being charted with inverse kinematics techniques involving rare isotope beams \cite{Tanihata2013}.
 The neutron drip line nucleus ${_4^{11}}$Be$_{7}$  provides a rare and precious window into the virtual processes
clothing neutrons which, through a rather conspicuous Lamb-shift-like effect, 
of the order of 10\% of the value of the Fermi energy, invert parity, thus allowing  for the emergence of a new magic number
diverse from that of Meyer-Jensen and extending, in the process, the limits of stability of nuclear species.

\begin{figure*}[t!]
\includegraphics[width=1.8\columnwidth]{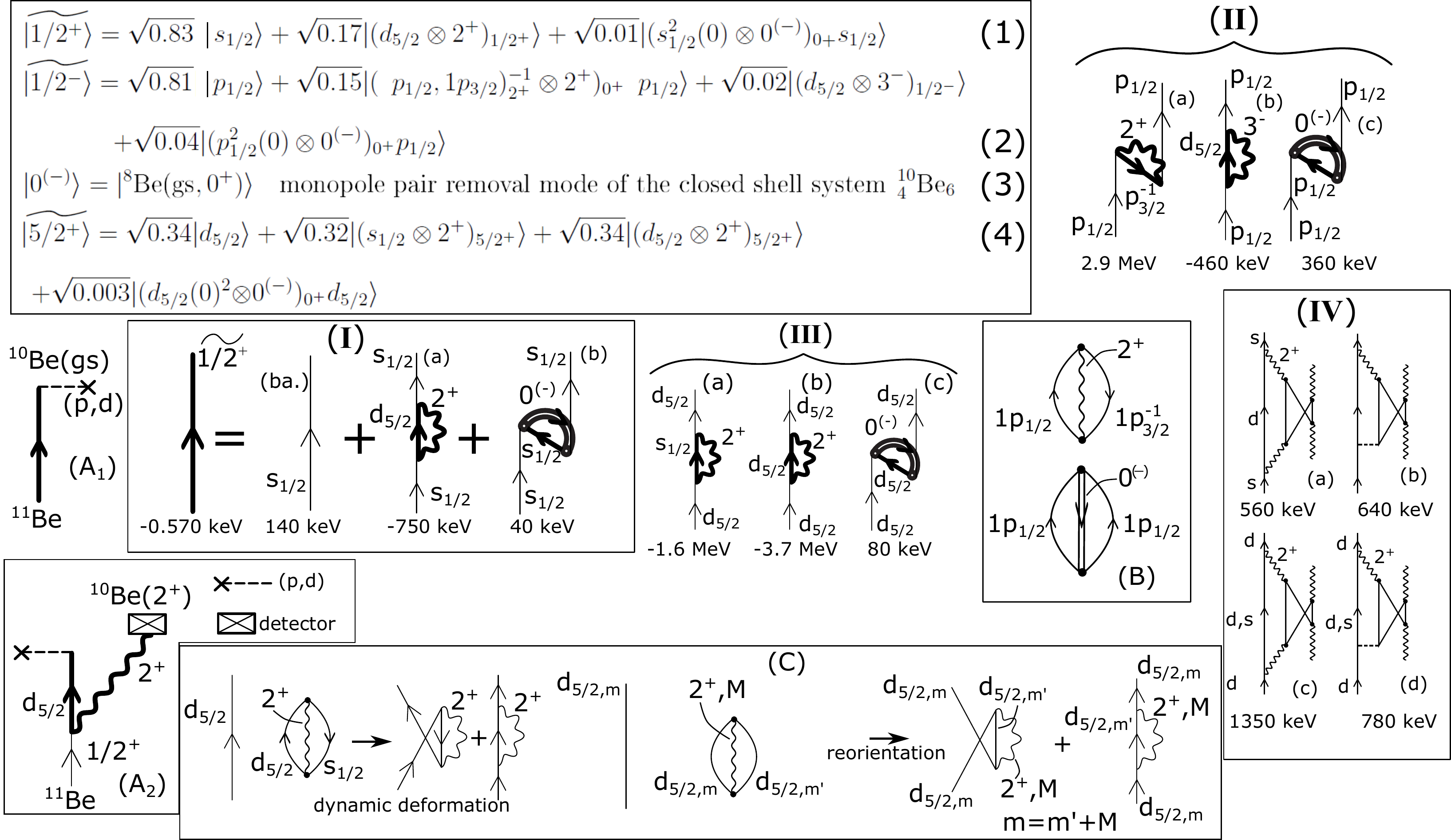}

\caption{(Upper box)  Wavefunctions associated with the renormalised 
single-particle levels labeled NFT in Fig. 1. (Lower Box) NFT diagrams
describing the processes responsible for the variety of components of the clothed states. Single-arrowed lines pointing up (down) describe
particle (hole) states, while wavy lines represent collective particle-hole (ph)vibrational states.Double arrowed lines pointing down describe
the correlated (hh) pair removal vibrational states.  
The dashed horizontal line describes the two-body multipole, separable interaction. In calculating 
the intermediate states
of the diagrams displayed in I-III the experimental and/or renormalised fermions bosons modes have been used (empirical renormalisation). This
is the reason why they are displayed in terms of bold face lines  and curves. The diagrams shown in IV take care of the Pauli principle violation 
of the two phonon states appearing in these intermediate states.  The label (ba.) in (I) stands for bare.
} 
\label{fig_diagrams}
\end{figure*}
 
Information about the nucleus $^{11}$Be 
is available regarding the $1/2^+,1/2^-$
and $5/2^+$ low-lying levels,  including the line shape  of the  $5/2^+$ resonance, in particular
from  inelastic and transfer reactions \cite{Iwasaki_a,Au.70,Zwieglinski,Fortier,Winfield,Schmitt} (Fig. 1(exp)). 
Analyses of these reactions have been  reported in \cite{Timofeyuk,Keeley,Deltuva1,Deltuva2,Lay,Dediego}. 
Information concerning the charge radius \cite{Nortershauser} and the dipole transition between the ground state and the first excited state
\cite{Kwan} is also available. Theoretical work 
on the structure of $^{11}$Be has been carried out,   
starting from parity
inversion between the 1/2+ and the 1/2- levels,
within the shell model \cite{Talmi},
while results of studies  based on the coupling of particles to vibrations and rotations are
found in refs. 
\cite{Otsuka}-\cite{Gori}   and  \cite{Dediego,Nunes,Fossez} respectively.
A static description of $^{11}$Be 
based on a deformed mean field was presented in \cite{Hamamoto}. 
Results of investigations 
within the framework of antisymmetrized  molecular dynamics were reported in \cite{Kanada}.
The outcomes of ab initio approaches  are  found in   \cite{Calci}.
Results of calculations of charge radii based on fermionic molecular dynamics have been reported in ref. \cite{Krieger}.

In this letter  we are able to achieve a unified and accurate description of both  structure and reaction aspects of 
$^{11}$Be. To do so, one has  to deal  simultaneously with 
the $p_{3/2}, p_{1/2},s_{1/2}$ and $d_{5/2}$ states, treating their interweaving 
with quadrupole and octupole (p,h) vibrations  and monopole 
pair vibrations on equal footing, as well as to take into account the mixing between  bound and continuum states.  
Furthermore, one has  to consider the  Pauli principle acting  between the particle explicitly considered and those participating in the vibrations as well as
between those participating
in two-phonon states. Because of
spatial quantization, the above requirement  involves both energies and single-particle radial  wave functions,  in particular  that of the 
$d_{5/2}$ resonance. The variety
of many-body clothing processes lead to  
important modifications of these radial wave functions, and thus of the corresponding one-particle transfer
form factors and escape particle wave functions, accounting for up to 50\% changes in the value of the one-nucleon transfer absolute cross
sections and  of the $5/2^+$ resonance decay width, in overall agreement with experimental data.  
It will be furthermore demonstrated that crucial information concerning the nature of the $5/2^+$ resonance and the role of
the quadrupole mode in dressing the nucleons moving around the Fermi surface is provided by
the reactions $^{10}$Be(d,p)$^{11}$Be(5/2$^+$, 1.833 MeV)  and $^{11}$Be(p,d)$^{10}$Be(2$^+$; 3.33 MeV)
which forces, in this last case, a virtual state, to become observable. A fact that aside from shedding light 
on retardation in clothing processes,
implies that 
particle-vibration coupled intermediate states which cloth the single-particle states have to be real,
empirical states concerning both energy and amplitude, as well as radial shape. Thus, $(NFT)_{ren}$ is not a calculational
ansatz but a quantal requirement. Within this context, it is of notice that self consistency within (NFT)$_{ren}$ implies that the final  results
$ \tilde \epsilon_j^{(i)}$ and $\tilde \phi_j(r)^{(i)}$ 
reproduce  the empirical input used for the intermediate states, while initial states (energies and wave functions) 
of the variety of graphical contributions
are  solutions of the bare potential. In other words, for each value of $\tilde \epsilon_j$ there can exist more than one radial 
function, depending on whether the nucleon  is moving around the ground state ( $i = gs$) or around an excited state 
of the core ($i = coll$) respectively.  Technically, $\tilde \phi_j (r)^{(i)}$ are the form factors associated with stripping and pickup reactions
around closed shells. 
For simplicity  we will drop the superscript $i$ in what follows.

\begin{figure*}
\includegraphics[width=0.5\columnwidth]{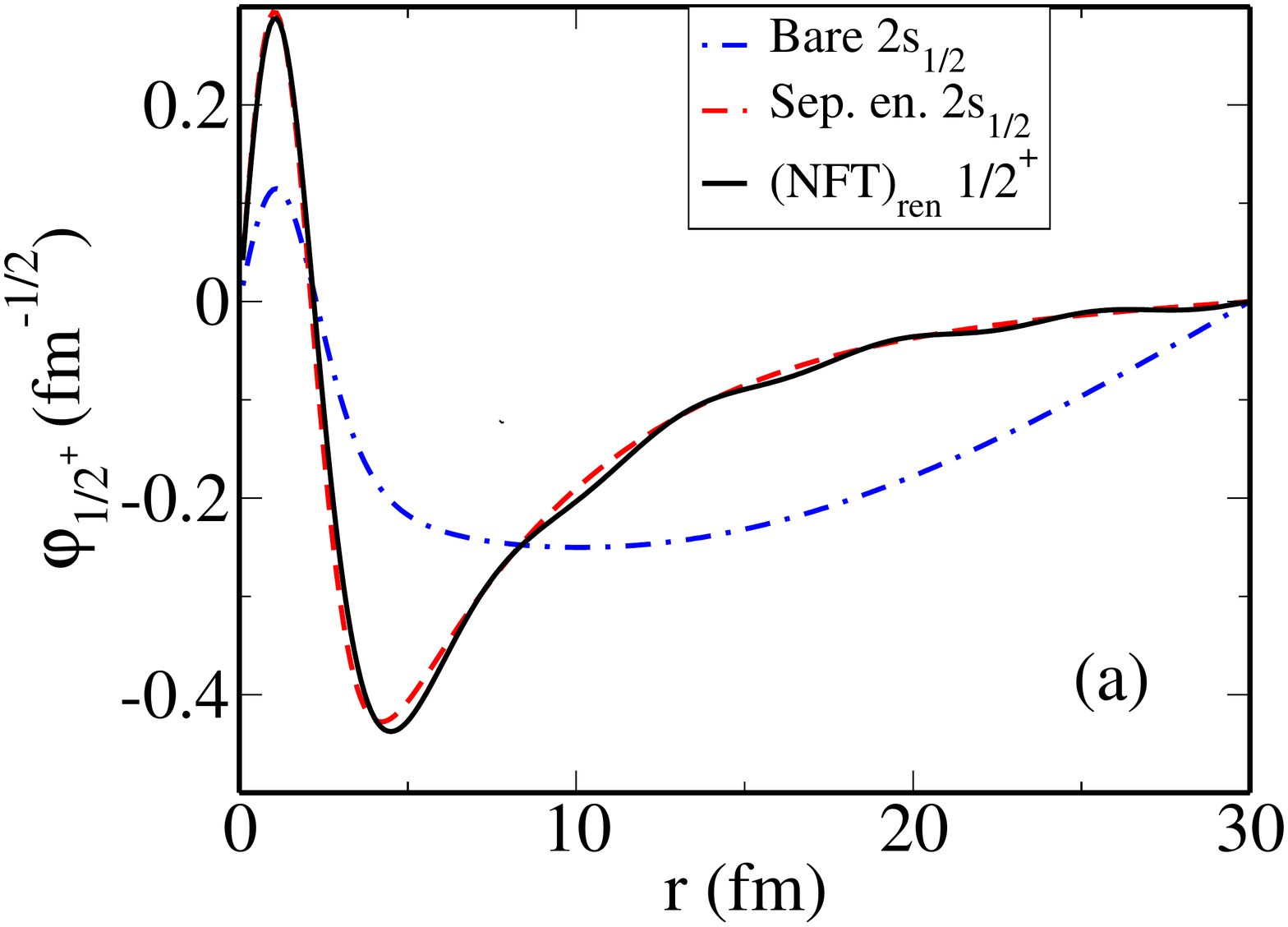}
\includegraphics[width=0.5\columnwidth]{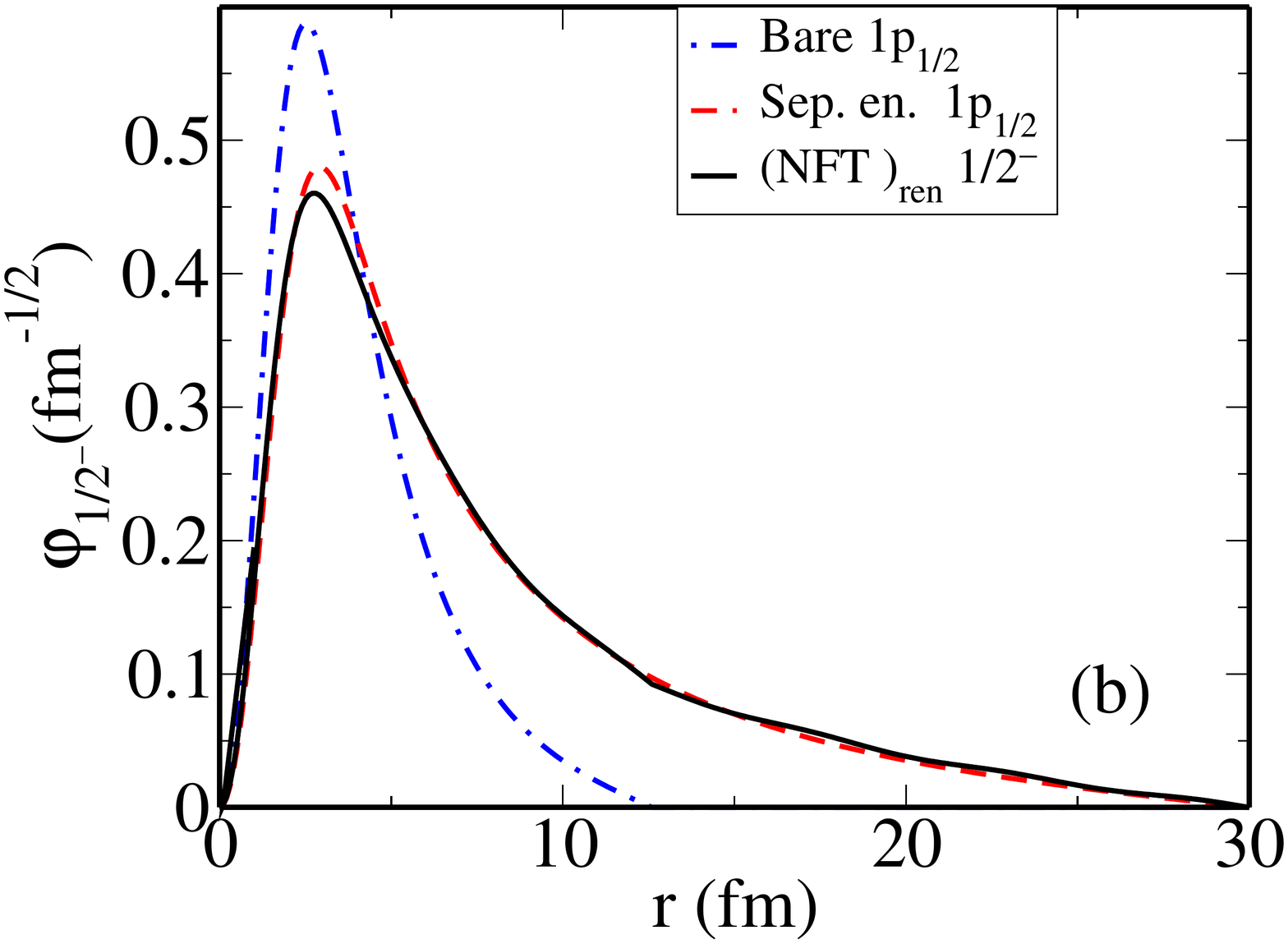}
\includegraphics[width=0.5\columnwidth]{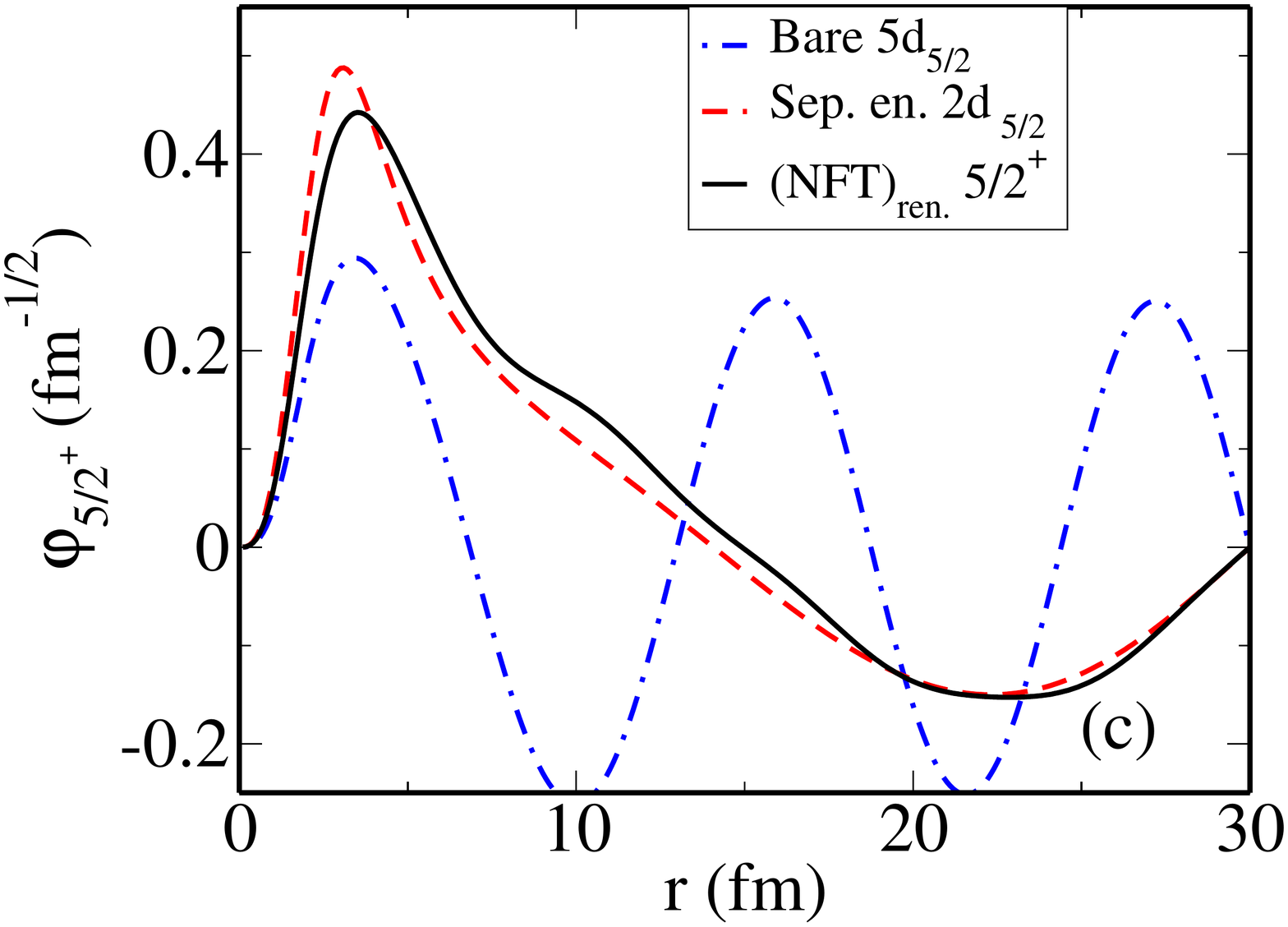}
\includegraphics[width=0.5\columnwidth]{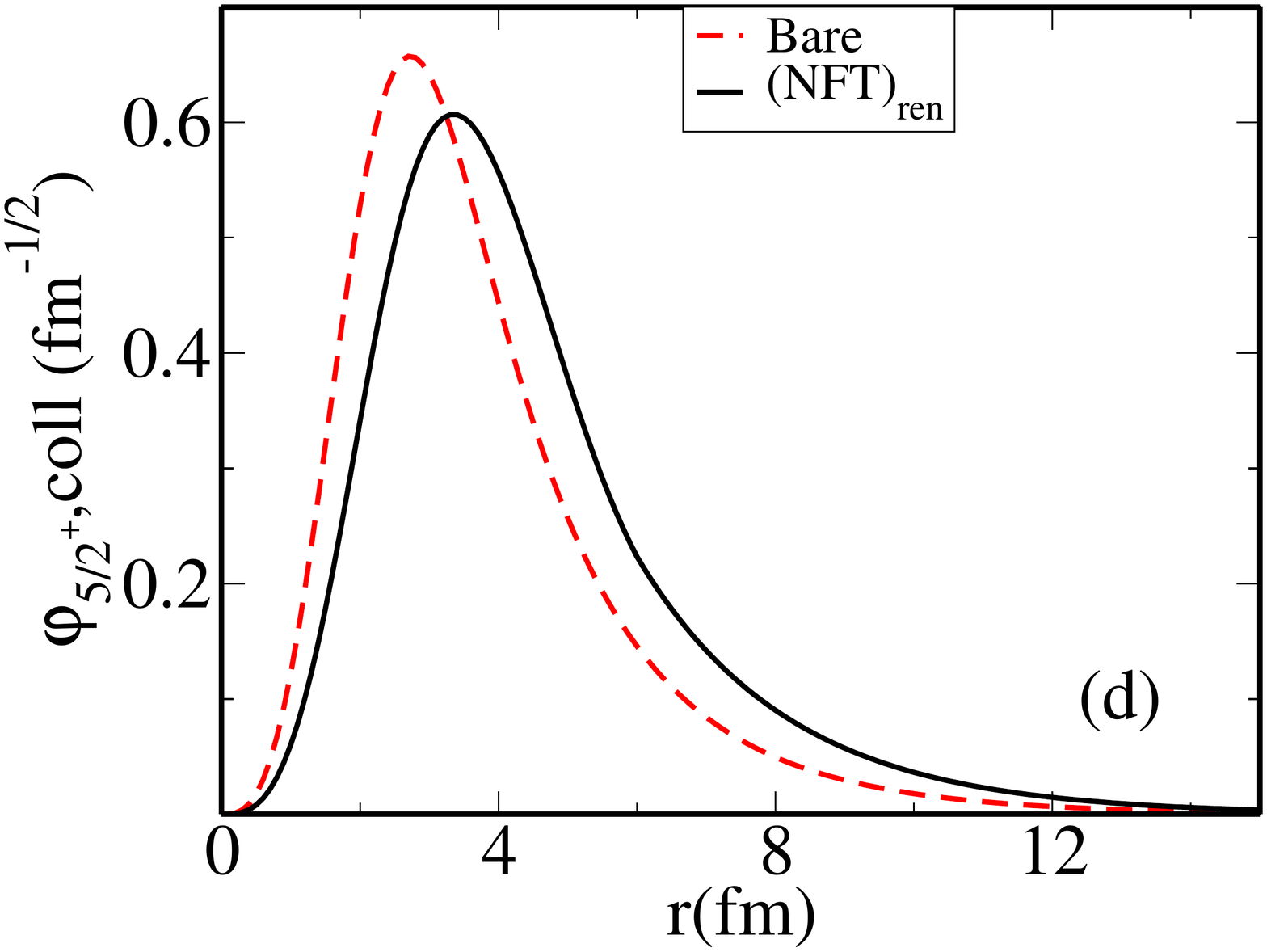}
\caption{ The form factors  of the $\widetilde {1/2^+} {\bf (a)}, \widetilde{1/2^-} {\bf (b)}, \widetilde{5/2^+} {\bf (c)}$ 
states  and ${\bf (d)}$ the form factor associated with the reaction
$^{11}$Be(p,d)$^{10}$Be(2$^+$)
calculated within the framework of NFT and  empirical normalisation
in a box of radius $R_{box}$ = 30 fm, divided by the associated amplitudes ($a_{\widetilde{1/2^+}} = \sqrt{0.83},
a_{\widetilde{1/2^-}} = \sqrt{0.81}, a_{\widetilde {5/2^+}} = \sqrt{0.34}, a_{(d_{5/2}\otimes 2^+)_{1/2^+}} = \sqrt{0.17}$, see Fig. \ref{fig_diagrams}, upper box)
are compared to the corresponding separation energy approximation
wavefunctions   and to the wave functions calculated with the bare potential parameterized as reported in Table \ref{table2}.}
\label{fig_waves}
\end{figure*}

\begin{figure}[h!]
 \includegraphics[width=0.9\columnwidth]{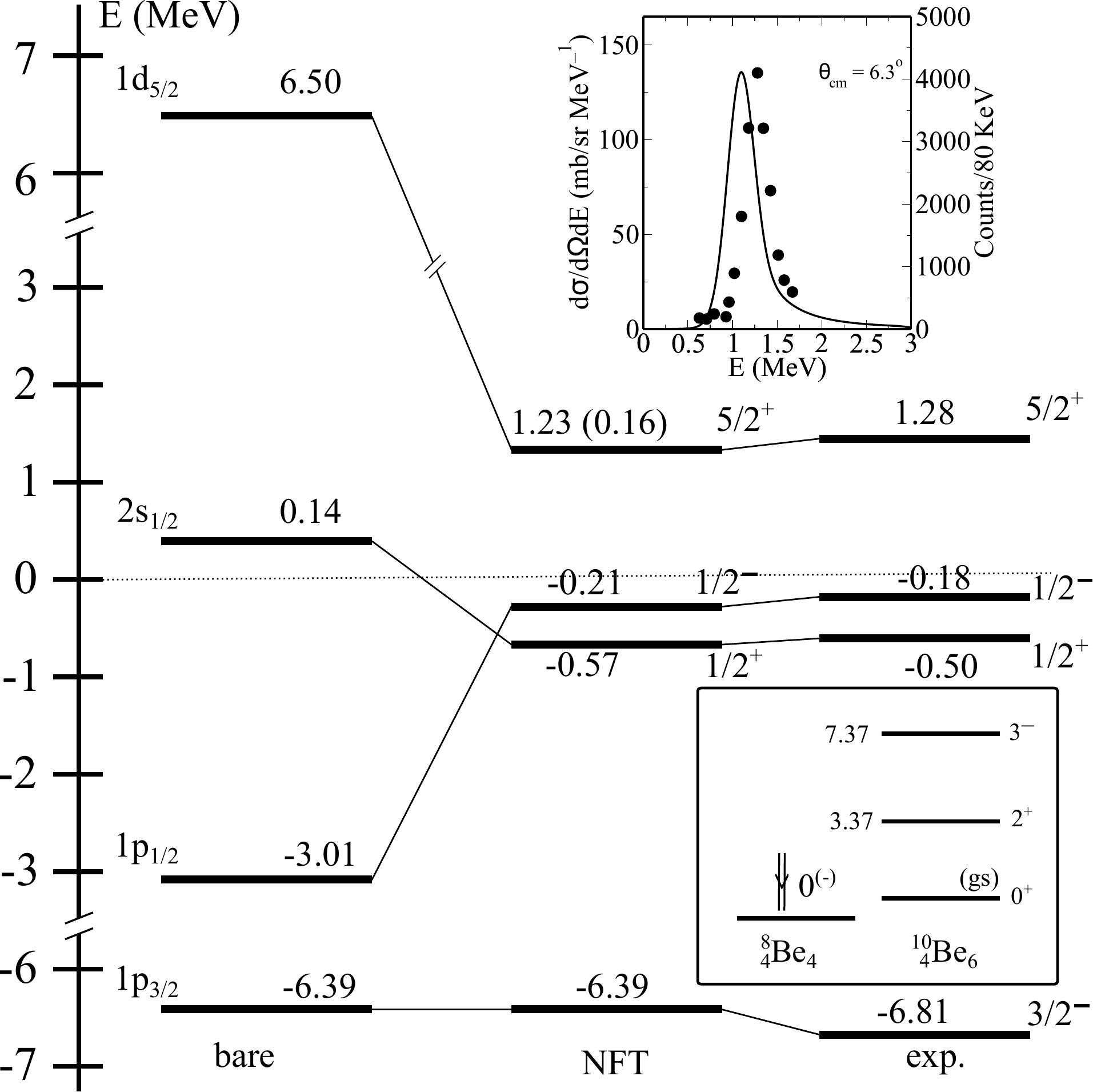}
\caption{ Low-lying spectrum of $^{11}$Be: (bare) unperturbed single-particle levels, 
solution of the bare mean field;
(NFT) renormalised levels; 
(Exp.)  experimental values.
The number on each thick horizontal line 
is the energy of the state in MeV.
The number in brackets correspond to the width of the $5/2^+$ resonance derived from the calculated elastic phase shifts. 
No phase shift data exist for this state, the 100 keV width often quoted being 
extracted from the $^9$Be(t,p)$^{11}$Be($5/2^+$)  reaction (\cite{Kelley}, Table 11.5).
The line shape of the $5/2^+$ resonance is displayed in the upper rhs corner in comparison with the data.
In the lower right corner inset,  the collective modes employed 
in the clothing of the bare single-particle  states (first column) are shown. 
The label $0^{(-)}$ indicates 
the correlated two-hole state, monopole ($J^{\pi} = 0^+$) pair removal mode.   
}
\label{fig_spectrum}
\end{figure}

Making use of this theoretical framework  (\cite{physscripta}  and  refs. therein, see also \cite{Mattuck,Idini2015})
we have calculated the variety of self energy diagrams, renormalizing selfconsistently  the motion 
of the odd neutron of $^{11}$Be in both configuration- (Fig. \ref{fig_diagrams}) and conformation 3D-space (Fig. \ref{fig_waves}).
The energies $\tilde \epsilon_j$ of the associated renormalised single-particle states (drawn with bold face arrowed lines 
in Fig. \ref{fig_diagrams}) are shown in Fig. \ref{fig_spectrum} (NFT) in comparison with the data (exp.), while the corresponding 
wave functions $\tilde \phi_j (r)$ are displayed in Fig. \ref{fig_waves} in comparison with those corresponding
to the separation energy approximation obtained  by adjusting the depth
of a standard Saxon-Woods potential   (\cite{BMI:69}, Eq. (2-180))
so as to reproduce the separation energy of each state in question making use of an effective mass
equal to the free nucleon mass. 
The quadrupole-
and octupole - as well as monopole pair removal - vibrational modes of the core $^{10}$Be (Fig. 3 inset)
and associated particle vibration
coupling (PVC) vertices were worked out in QRPA ($\beta_2^n$ =0.9 and $\beta_3^n$ = 0.28) and in RPA respectively 
with the help of separable interactions of self consistent strength.
They are drawn
in Fig. 1 in terms of bold face wavy curves (quadrupole and octupole) and arrowed double lines (pair mode).
We remark that the error in the self-energy associated with  overcounting  (bubble correction)
in the sum over  intermediate particle-phonon configurations,  amounts to  at most to  a few percent effect in the present case  (Suppl. Mat., Section 3) . 

The bare states (thin arrowed lines in Fig.1) were determined as solutions of a mean field potential 
of radial dependent $k-$mass. This field 
was parameterized as a Saxon-Woods potential  
with a spatially dependent effective mass of
value $m_k(0) =0.7 m$ at the 
center of the nucleus and  $\mu = 0.91 m$ (reduced mass value) far from it.
 The parameters were varied starting from values 
reproducing the shape  of the mean field obtained with the  SGII Skyrme interaction (Table 1; see also Suppl. Mat, Section 1).
The resulting  bare energies  $\epsilon _j$  and wave functions $\phi_j$ are shown in Figs. \ref{fig_spectrum} (bare)
and \ref{fig_waves} respectively.  

The clothed states associated with a given quantum number result from the iterative diagonalization 
of the PVC Hamiltonian in a space  composed of single-particle and of particle-phonon states,
making use of self-energy function techniques (Fig. \ref{fig_diagrams}(I), see also Suppl. Mat., Section 2).
We included in the calculation the $s_{1/2},p_{1/2}$ and $d_{5/2}$ single-particle states up to an energy $E_{cut}$ = 25 MeV,
imposing vanishing boundary conditions at $R= $ 30 fm (continuum discretization),
leading to  matrices of typical dimension $N \approx 300$.
The lowest $p_{3/2}$ state was also included. 
The diagonalization process is equivalent to the solution of a system of coupled integro-differential equations (see   \cite{Pinkston}  and Suppl. mat., Section  7).
The interference between the amplitudes 
of the  various single-particle  states determines the radial dependence of the single-particle component of the eigenstates, in particular in
the surface region (Fig. \ref{fig_waves}).

The most conspicuous effect emerging from these 
NFT results is the (absolute value) reduction
 of the energy  difference between positive and negative parity states (Fig. \ref{fig_spectrum}):
 a factor of ten 
in the case of $1/2^-,1/2^+$ states (from $\epsilon_{1/2^+} - \epsilon_{1/2^-} $ = +3.2  MeV (bare)
to  $\widetilde{\epsilon_{1/2^+}} - \widetilde{\epsilon_{1/2^-}} = $   -0.3  MeV (NFT)) and of six
in the case of $5/2^+, 1/2^-$ states 
( from $\epsilon_{5/2^+} - \epsilon_{1/2^-} $ = +9.5 MeV (bare)
to  $\widetilde{\epsilon_{5/2^+}} - \widetilde{\epsilon_{1/2^-}}= $   + 1.5 MeV (NFT)).
While parity inversion is only observed in the first case,
the second had a close call, playing an essential role in the mechanism which is
at the basis of ($1/2^+,1/2^-$) parity inversion ($1/2^+$ self energy polarisation (PO) 
and $1/2^-$ correlation (CO) processes \cite{Mahaux} arising from Pauli principle of the odd
neutron and ZPF of the core ground (vacuum) state).  
 It is an open question whether parity inversion between $1/2^-$ and $5/2^+$ states can ever be observed in nuclei. 

As testified by the components of the clothed states displayed in Fig. \ref{fig_diagrams} (upper box), the physics at the basis of 
the  results reported above  is mainly related to the coupling of single-particle to  
dynamical, $\omega-$dependent, quadrupole deformation,  conspicuously 
modified by Pauli principle corrections.
Let us start with the $5/2^+$ resonance. This state  is prone to acquire a dynamical quadrupole moment (reorientation
effect, see Fig. \ref{fig_diagrams}  inset (C)). This is because 
the particle-vibration coupling  of the $d_{5/2^+}$ with itself through the excitation of the quadrupole vibration 
of $^{10}$Be, involving a rather confined single-particle resonant state radial wave function (Fig. \ref{fig_waves} (c)) ,results in a large 
value of  $<\tilde \phi_{5/2^+} | R_0 dU/dr | \tilde \phi_{5/2^+}>$ and thus in a large non spin-flip  
PVC matrix element 

$(<5/2^+ \otimes 2^+)_{5/2^+}|H_c|  d_{5/2}> \approx$ -4.6 MeV) leading to an amplitude of $\sqrt{0.34}$ 
for the many-body state $|(d_{5/2} \otimes 2^+)_{5/2^+}>$
(Fig. \ref{fig_diagrams} , upper box, Eq.(4)). 
An equally   important component ($\approx \sqrt{0.34}$) 
is associated with the coupling of  the $5/2^+$ state to the $s_{1/2}$ state, again through the quadrupole mode.
This coupling allows the bare $d_{5/2}$  resonance ($\epsilon_{5/2^+} \approx$ 6.5 MeV, Fig. \ref{fig_spectrum}), 
to explore halo-like regions of the system
and to lower its   kinetic energy.  
Together with the effects of the couplings discussed above and the repulsive one
associated with the coupling to the pair removal mode,  it results in an overall  energy decrease of the
$5/2^+$ strength and in the buildup of a narrow  (Fig. \ref{fig_spectrum} inset line shape) resonance
with centroid at $E = 1.23 $ MeV and a width of 160 keV calculated from elastic scattering phase shifts (Suppl. Mat., Sect. 2). 
In turn, the $2s_{1/2}$ wave function mixes with the ($d_{5/2} \otimes 2^+$) configuration and acquires
a component of sizable amplitude ($\approx \sqrt{0.17}$)  
lowering, in the process, its energy by about 750 keV (Fig. \ref{fig_diagrams} (I)(a)), 
the repulsion due to Pauli principle
and arising from the mixing with the ($s_{1/2} \otimes 0^{(-)}$) configuration 
being rather modest (Fig. \ref{fig_diagrams} (I(b))). In other words, 
the $5/2^+$ state plays, through its coupling to the $2^+$ vibration of the core,  an essential role in the 
$(1/2^-,1/2^+)$ parity inversion phenomenon.

The zero-point fluctuations (ZPF) associated with the $^{10}$Be core, both those arising form  the quadrupole vibrations as well
as from the monopole pair removal mode  (boxed inset (B) of Fig. \ref{fig_diagrams} ) make virtual use
of the $p_{1/2}$ single-particle state. The first and third NFT diagrams displayed in Fig. \ref{fig_diagrams} (II) properly treat
the problem of the identity  of the particles appearing explicitly and the particles participating in the collective motion. As a result,
the phase space of the clothed $1p_{1/2}$ state becomes smaller than the bare one, its binding  
becoming weaker by about 3 MeV
due to the Lamb-shift-like process shown in Fig. \ref{fig_diagrams} (II)(a) and by 360 keV due to that displayed in 
Fig. \ref{fig_diagrams}(II)(c), the contribution
of the second graph (Fig. \ref{fig_diagrams} (II)(b)) being attractive. 
The numbers quoted above contain, among other things, the renormalisation contribution of the Pauli correcting diagrams 
shown in Fig. \ref{fig_diagrams} (inset(IV) right hand side) 
and associated with the  implicit presence of two-phonon states in intermediate, virtual configurations, drawn in bold face.

The radial dependence of the many-body wavefunctions and the phonon admixture can be probed by the one-neutron transfer reactions 
$^{10}$Be(d,p)$^{11}$Be and $^{11}$Be(p,d)$^{10}$Be(2$^+$), 
populating the low-lying $1/2^+,1/2^-$ and $5/2^+$  states of $^{11}$Be and  the first 
excited $2^+$ state of $^{10}$Be, and proceeding 
through the form factors displayed in Figs. \ref{fig_waves}(a)-(c)   and Fig. \ref{fig_waves}(d) respectively.  
Concerning the latter, we remark that the asymptotic decay constant of the
 $d_{5/2}$ radial wave function
associated with the $2^+ \otimes d_{5/2}$ configuration admixed in the $1/2^+$ ground state of $^{11}$Be 
displays a binding energy 
$\tilde \epsilon_{1/2^+} - \hbar \omega_{2^+} =$ -3.8 MeV. 
It is a natural outcome of (NFT)$_{ren}$  to give rise, through PVC and Pauli mechanism, to the proper clothing 
of the $d_{5/2}$ orbital so as to be able "to exist" inside the $|\widetilde{s_{1/2}}>$ state  as a virtual, intermediate configuration
(Fig. \ref{fig_diagrams}(I)(a) and IV(a) and (b)). The associated asymptotic $r-$behaviour
results from the coherent superposition
of many  continuum states,  and its spatial dependence in the surface region  (Fig. \ref{fig_waves} (d)) can hardly be simulated 
by making use of  a bound single-particle wavefunction of a properly parameterized
single-particle potential (separation energy approximation), in agreement with the result of previous studies \cite{Winfield}.

\begin{figure*}
\includegraphics[width=0.5\columnwidth]{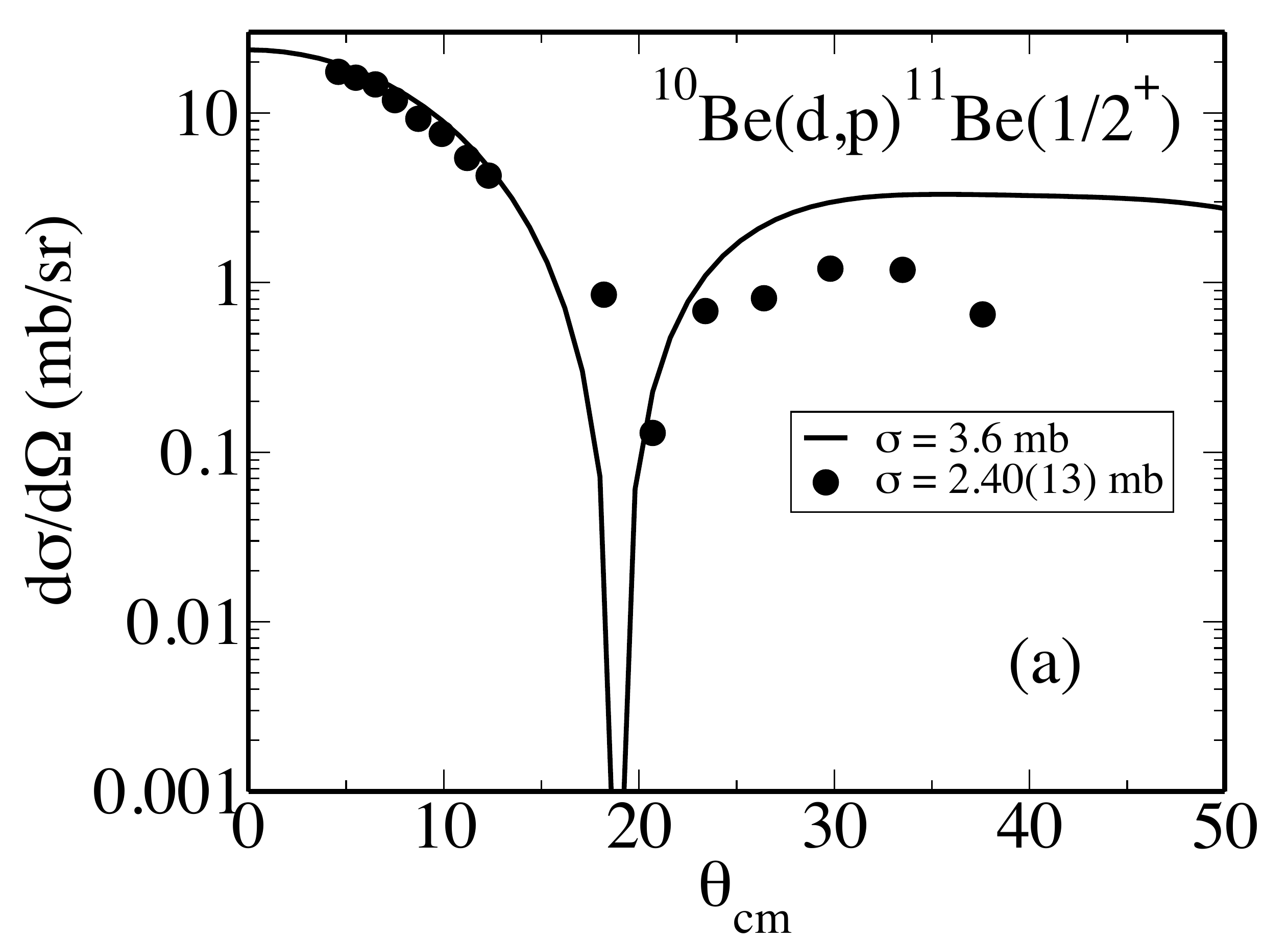}
\includegraphics[width=0.5\columnwidth]{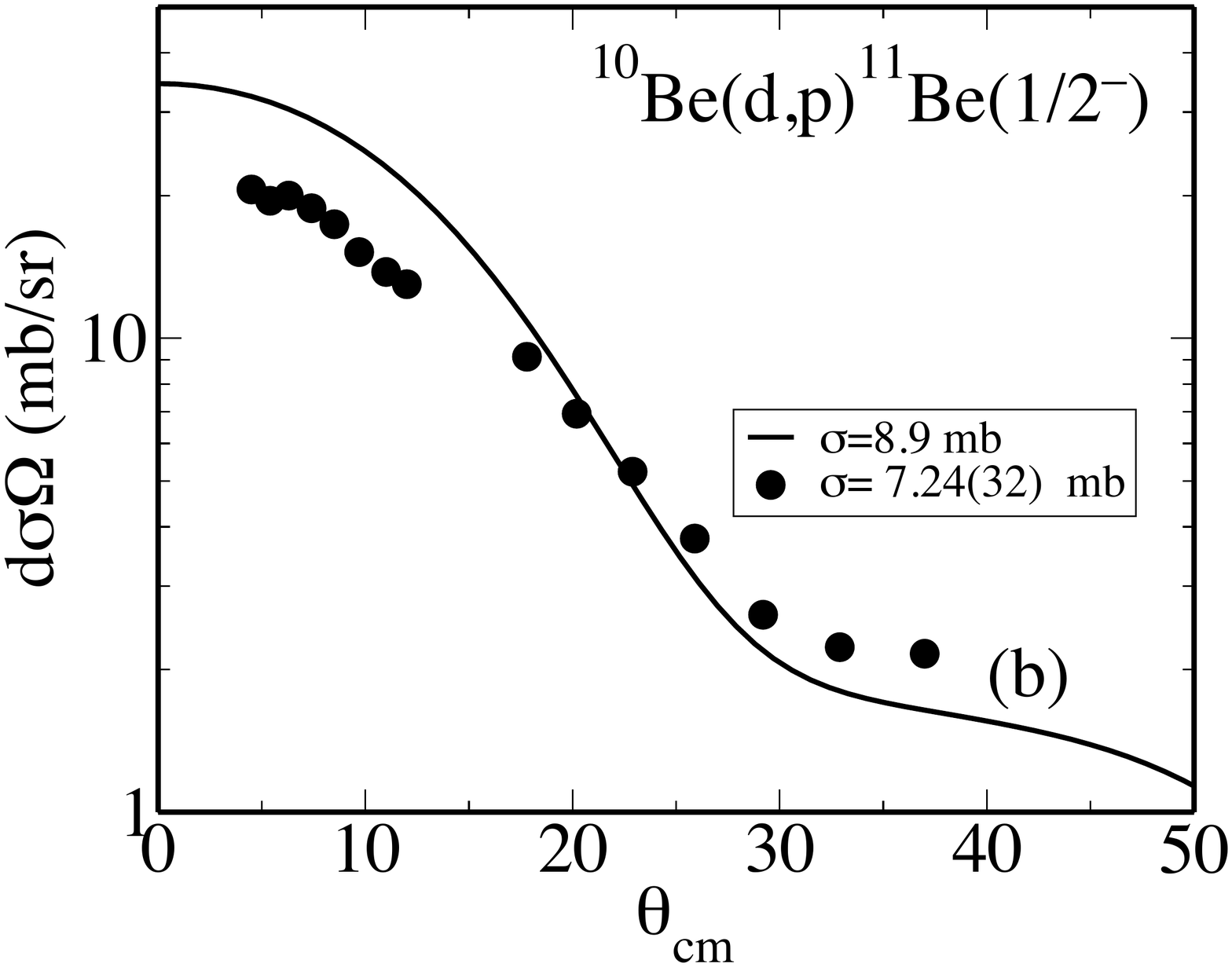}
\includegraphics[width=0.5\columnwidth]{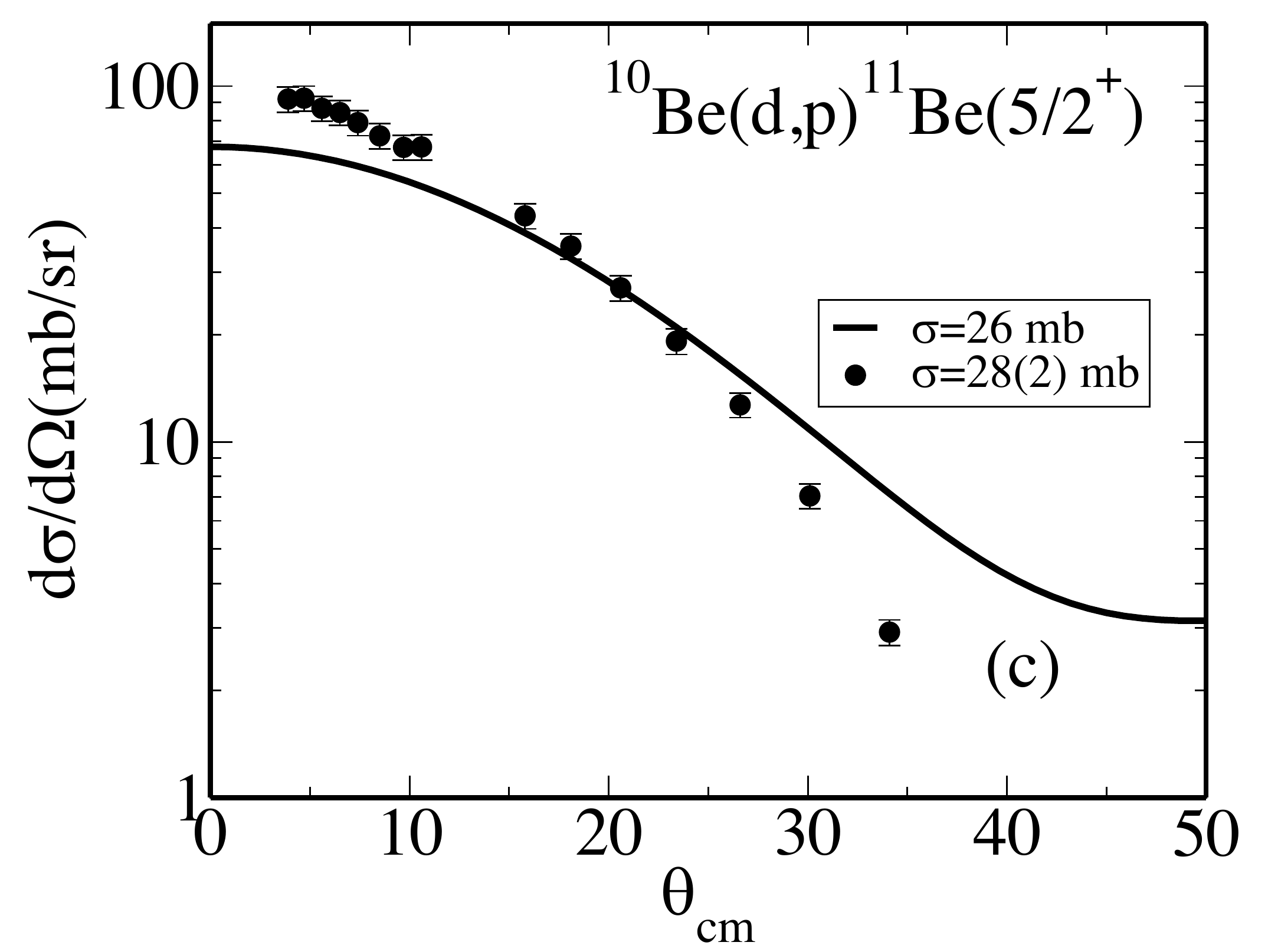} 
\includegraphics[width=0.5\columnwidth]{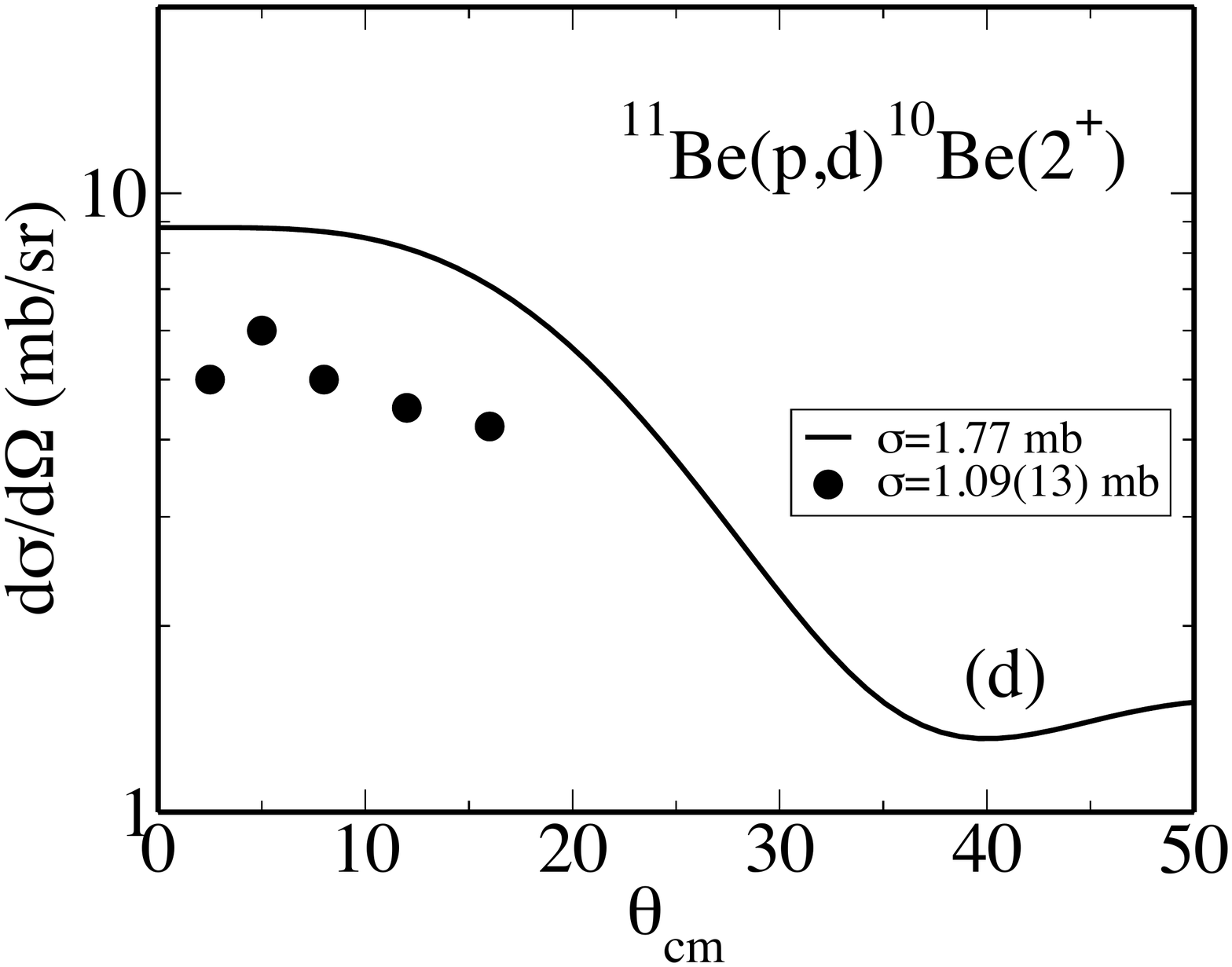}
\caption{ {\bf (a-c)} (continuous curve) Absolute differential and summed cross sections associated with the reactions  
$^{10}$Be(d,p)$^{11}$Be, populating the   ${1/2^+, 1/2^-}$, and ${5/2^+ }$ states ($E_d = $ 21 MeV).
The experimental data  \cite{Schmitt}  are displayed in terms of solid dots.
{\bf (d)} Same as before, but for the reaction   $^{11}$Be(p,d)$^{10}$Be, populating the  ${2^+ }$ state ($E_p = $ 35.3 MeV/n)
\cite{Winfield}. }
\label{fig_cross}
\end{figure*}

The (NFT)$_{ren}$  form factors shown in Fig. \ref{fig_waves} were used,  
together with  global optical  potentials \cite{opt_pot1,opt_pot2}, to calculate the one-nucleon stripping and pickup absolute differential 
cross sections of the reactions mentioned above. The results provide an overall account  of the experimental 
findings  (Fig. \ref{fig_cross}). Within this context we remark that  the pickup process shown in inset
(A$_1$) of Fig. \ref{fig_diagrams} and populating $^{10}$Be ground state  implies the action of
the external (p,d) field on the left hand side (lhs) of the graphical representation of Dyson equation shown in Fig, 
\ref{fig_diagrams}(I), and involves, at the same time, the use of the corresponding radial wavefunction
as formfactor (Fig. \ref{fig_waves}(a)). 
In the case of the population of the first $2^+$ excited state of $^{10}$Be (inset $A_2$), the (p,d) field acts on the 
$(d_{5/2} \otimes 2^+)_{1/2^+}$ virtual state of the second graph of the right hand side (rhs)  of this equation
(Fig. \ref{fig_diagrams}(I)(a)), involving this time the form factor shown in Fig. \ref{fig_waves}(d). Summing up, insets 
(A$_1$) and (A$_2$)  and diagrams (I) of Fig. \ref{fig_diagrams} testify to the subtle effects resulting  from the unification of (NFT)$_{ren}$ of structure 
and reactions discussed in \cite{physscripta}, and operative in the cross sections shown in Fig. \ref{fig_cross}, as a result of the 
simultaneous and self consistent treatment of configuration and 3D-space.  Within this context
the bold face drawn state $|d_{5/2} \otimes 2^+)_{1/2^+}>$ shown in Fig. \ref{fig_diagrams}(I)(a) 
and the wave function displayed in Fig. \ref{fig_waves}(d) can be viewed as {\it on par} structure and reaction 
intermediate elements of the quantal process $^{11}$Be(1/2$^+$)(p,d) $^{10}$Be($2^+$).

Particle-vibration coupling leads to important renormalization effects 
in the radius and in the dipole electromagnetic transitions of the system,
the two phenomena being closely related. This is due to the poor overlap between the resulting 
renormalised halo radial wave functions and those of the core nucleons which screen 
the symmetry potential, impeding the GDR to shift the $1/2^+ \to 1/2^-$ single-particle 
E1-strength to high energies in the attempt to exhaust the EWSR \cite{Hamamoto,physscripta,Bortignonbracco}
.

The strength of the dipole  transition connecting the ground  and the first excited states was calculated and compared to the experimental 
value $B(E1) = 0.102 \pm$ 0.002 e$^2$ fm$^2$. 
Estimates carried out  using  single-particle wavefunctions corresponding
to the separation energy approximation and normalised to one,  give
$B(E1; I_i \to I_f)=  \frac{e^2_{eff}}{2\pi} \frac{| <I_f|| i  M(E1)|| I_1>|^2}{2I_1+1} = 0.29 {\rm e^2 fm^2}$,
with  $e_{eff}$ = 4/11.  
Including the renormalizations associated  with  the phonon admixture of 
the  $1/2^+$ and $1/2^-$ state leads to $B(E1; 1/2^- \to 1/2^+) $ = 0.11 e$^2$ fm$^2$,  
that is to a reduction of over a factor 
of 2  bringing theory in overall agreement with  the data (cf. Suppl. Mat., Section  5).


 \begin{table}[h!]
\begin{center}
\begin{adjustbox}{max width=\textwidth}
\begin{tabular}{| c |  c | c | c |}
\hline 
$ V$ (MeV) & $V_{ls}$ (MeV) &
$a$ (fm)  &  $R$ (fm)   \\ \hline
 70 & 14 & 0.81& 2.10 \\ \hline 
\end{tabular}
\end{adjustbox}
\end{center}
\caption{
 Optimal values of  
the depth, radius, diffuseness and spin-orbit strength 
of the bare mean field potential $V$,$V_{ls}$,$a,R$ (see \cite{BMI:69} Eq. (2-180)).
 } 
\label{table2}
\end{table}

Let us now discuss the isotopic shift of the charge radius (Suppl. Mat., Section 6).
The corrections to the charge mean square radius $<r^2>_{^{10}Be}$ arising
from the addition of a neutron are obtained applying the external field operator $r^2$ to the 
different particle and collective  vibration lines and curves of the diagrams appearing in the rhs
of the graphical equation displayed in Fig. \ref{fig_diagrams}(I). The summed (recoil) contributions 
associated with the $s_{1/2}$ state  (graphs (b) and (a))   $\left( <r^2>^{1/2}_{1/2^+}/11 \right )^2
(<r^2>^{1/2}_{1/2^+} $ = 7fm) is to be multiplied by the square of the renormalised $\widetilde{|1/2^+>}$
state single-particle amplitude (=0.83, Fig. \ref{fig_diagrams}(1), upper box). Those associated 
with the $d_{5/2}$ and $2^+$ elementary modes appearing in the intermediate  state of graph
(a) of Fig. \ref{fig_diagrams}(I) lead to (recoil) ( $<r^2>^{1/2}_{5/2^+}/11)^2$ and (dynamical deformation)
($<r^2>_{5/2}^+ \beta_2^2/2\pi)$ contributions respectively,   and are to be multiplied by the square amplitude 
(=0.17) associated with the $(d_{5/2} \otimes 2^+)_{1/2^+}$ configuration (Fig. \ref{fig_diagrams}, upper box). 
Contributions arising from graph (b) of Fig. \ref{fig_diagrams}(I) are not considered in keeping with their small relative values.
The resulting theoretical prediction $<r^2>_{^{11}Be}^{1/2}$ = 2.48 fm, accurately 
reproduces the experimental finding 2.44 $\pm$ 0.06 fm.

We conclude   that a consistent fraction of the clothed  one-neutron
halo valence single-particle states $|\widetilde {j}>$ of $^{11}$Be, as much as 60\% in the case of the $|\widetilde{5/2^+}>$ state, corresponds to many-body configurations
resulting from Pauli principle and particle-vibration coupling effects.
These components, which modify both the single-particle content and the radial dependence
of the associated wave functions,
 play a central role in providing 
an overall quantitative account
of an essentially   "complete" set of data characterising  the structure of 
$^{11}$Be. 
The existence of a bare potential, empirically determined (minimisation procedure) 
 as an appropriate basis
for nuclear many-body studies, can be considered as a benchmark of mean field theories.

\section{Acknowledgment}
F.B and E.V. acknowledge  funding from
the European Union Horizon 2020 research and innovation
program under Grant Agreement No. 654002.

\clearpage

\setcounter{equation}{0}
\setcounter{figure}{0}

\section{SUPPLEMENTARY MATERIAL}

\subsection{The mean field }

The  bare mean field is taken to be  a Saxon-Woods potential containing a central and a spin-orbit term  as in [1]
(see Eq. (2-180)) :
\begin{equation}
U(r) = V_{WS} f(r) + V_{ls} (\vec l \cdot \vec s) r_0^2  \frac{d f(r)}{dr},
\end{equation}
with 
\begin{equation}
f(r) = \left[ 1 + {\rm exp} \left( \frac{r -R_{WS}}{a_{WS}} \right) \right]^{-1}.
\end{equation}

The $k-$mass has been parameterised according to,
\begin{equation}
m_k(r) = \mu_{red} - {C_m} \left[1 + {\rm exp} \left( {\frac{r - R_m}{a_m}}\right)  \right]^{-1},
\label{meff}
\end{equation}
with $a_m$ = 0.5 fm, $R_m = 2.34$ fm. Far from the nucleus,  the effective mass 
becomes equal to the reduced mass, $\mu_{red} = 10/11 = 0.91$.
 
\begin{figure}[h!]
\includegraphics[width=0.4\textwidth]{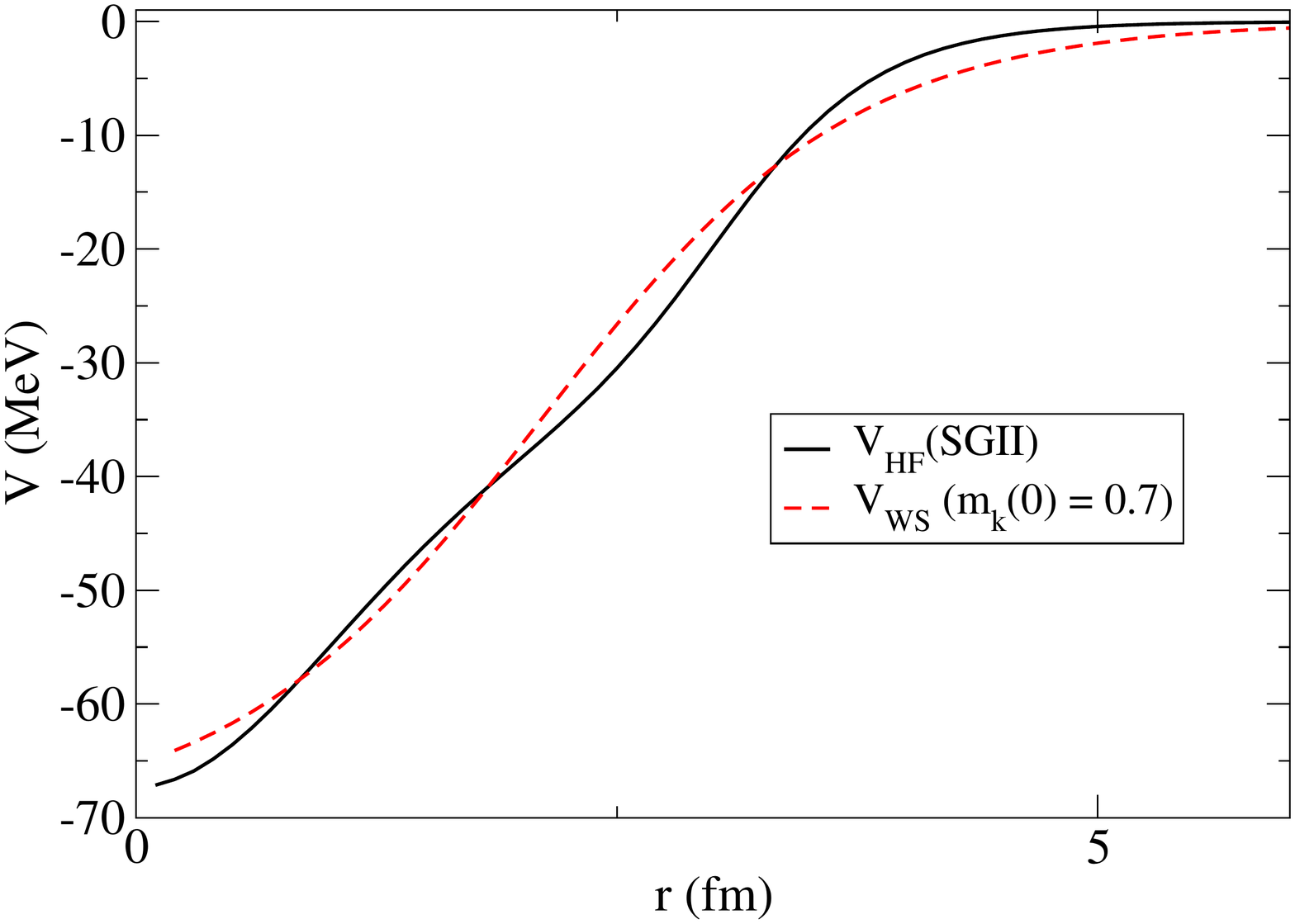}
\includegraphics[width=0.4\textwidth]{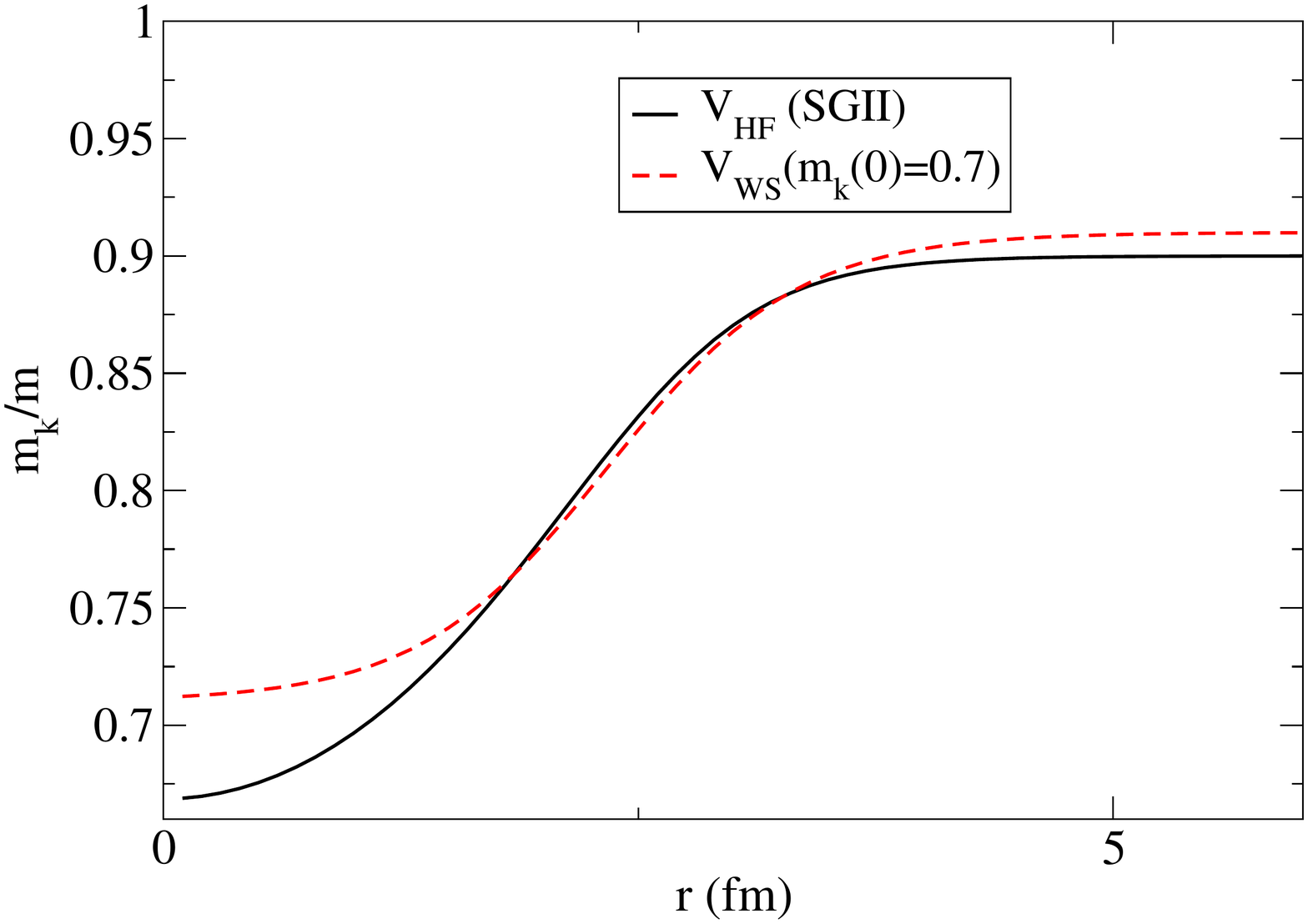}
\caption{The central potential (top panel) and the effective mass  (bottom panel) calculated with the fitted  Saxon-Woods potential
 are compared to the corresponding quantities 
obtained from a Hartree-Fock calculation with the SGII Skyrme interaction.}
\label{fig:sgii}
\end{figure}

The parameters of the  central potential, as well as the deformation parameters $\beta_2$ and
$\beta_3$  have been varied  so that the solution of the Dyson
equation in a box of radius $R_{box} = $ 40fm reproduces  the experimental energies after renormalization
(NFT)$_{res}$ clothing).
In other words, the deviation from theoretical output and experiment to be minimized  is 
\begin{eqnarray}
\chi^2 = (\tilde \epsilon_{s_{1/2}} - \epsilon^{exp}_{s_{1/2}})^2 + (\tilde \epsilon_{p_{1/2}} - \epsilon^{exp}_{p_{1/2}})^2
+ (\tilde \epsilon_{d_{5/2}} - \epsilon^{exp}_{d_{5/2}})^2 & \nonumber  \\
 +(0.15 \times ( \tilde \epsilon_{p_{3/2}} - \epsilon^{exp}_{p_{3/2}}))^2 +
(0.01 \times  (\beta_2 V_{WS} R_{WS} - 119))^2 ,
\end{eqnarray}
where we have also introduced the value of $\beta_2 V_{WS} R_{WS} = 119$ MeV fm derived from  the experimental analysis of proton inelastic scattering
used by  [2]
We have normalized the differences between theory and experiment, so that 
a similar relative error for the various quantities has a similar impact on the value of $\chi^2$. 
We have also required that the admixture of the $2^+$ phonon in the wave function of the dressed $1/2^+$, $1/2^-$  and $5/2^+$ states is larger than 10\%.
 The  central term of the  resulting Saxon -Woods potential and the associated effective mass are shown in Figs. \ref{fig:sgii}, where they
 are compared with the corresponding quantities obtained from a Hartree-Fock calculation  with the SGII interaction.
 
 \subsection{Calculation of the single-particle energies and spectroscopic factors}
 
 The matrix elements of the self-energy matrix $\Sigma_{i,k} (E)$ between
 a pair of single-particle states of $^{11}$Be labeled   $i \equiv \{n_i,l_i,j_i\}$ ,$ k \equiv \{n_k,l_k,j_k\}$  and of energies  $(\epsilon_i, \epsilon_k > \epsilon_F)$  
is written as, 
 \begin{equation}
 \Sigma_{i,k} (E) = \delta_{l_i,l_k}\delta_{j_i,j_k} [ \delta_{n_i,n_k} \epsilon_{n_i lj} + \Sigma^{pair}_{i,k}(E) + \Sigma^{surf}_{i,k}(E)],
  \end{equation}
  where the contribution associated with the coupling to surface vibrations is given by 
  \begin{equation}
\Sigma^{surf}_{i,k}(E) =  \sum_{\lambda,n,c, \epsilon_c > \epsilon_F}  \frac{h(i , c \lambda n) h(k, c \lambda n)}{E - (\epsilon_c + \hbar \omega_{\lambda,n})}
  +
   \sum_{\lambda,n,c, \epsilon_c < \epsilon_F}  \frac{h(i, c \lambda n) h(k, c \lambda n)}{E - (\epsilon_c - \hbar \omega_{\lambda,n})}
  \end{equation}
  and that associated with the coupling to pair vibration by 
  \begin{equation}
\Sigma^{pair}_{i,k}(E) =  \sum_{\lambda,n,c, \epsilon_c >\epsilon_F}  \frac{h(i , c 0^{-}) h(k, c 0^{(-)})}{E + (\epsilon_c - \hbar \omega^{rem})}.
  \end{equation}
The sum is taken over the phonons $\lambda ,n$ and over a set of intermediate states $c \equiv \{n_c l_c j_c \}$.
 The matrix elements connecting single-particle level $a \equiv \{nlj\}$ with an intermediate
 particle-phonon state $b \lambda$ (surface), or  $c0^{(-)}$ (pair)   are denoted by $h(i,c \lambda n) $ and $h(i,c0^{(-)})$

We include in the calculations particle states of angular momentum $s_{1/2},p_{1/2} $ and $d_{5/2}$   up to an energy cut off $E_{cut} = 25 $ MeV.
We include  a single hole state, namely the  $1p_{3/2}$ orbital.  The self-energy matrix is diagonalized separately for each 
 $l,j$, on an energy mesh,
and  the resulting eigenvalues $\tilde \epsilon_{lj}$ are a solution of the Dyson equation,
$ \sum_k [\Sigma_{ik} (\tilde \epsilon_{lj}) \times x^{lj}_k] = \tilde \epsilon_{lj} x^{lj}_i $.

Each solution  has a
single-particle component and a collective component.  Denoting the eigenvectors by $x^{lj}_i$ ( normalised so that $\sum_i (x^{lj}_i)^2 =1$)  ,
the single-particle component is written as  
\begin{equation} 
\tilde \phi_{lj} (r) = \sum_i x^{lj}_i \phi_{n_ilj} (r),
\end{equation}
while the collective part associated with particle-phonon configurations $c, \lambda n$ 
is given by  
\begin{equation}
\tilde \phi^{coll}_{\lambda, n ,l_c j_c } (r) =  \sum_{n_c} R^{coll}_{(c,\lambda n)_{lj}} \phi_{n_c l_c j_c} (r),
\end{equation}
where $ R^{coll}_{(c,\lambda n)_{lj}}= \sum_i x^{lj}_i  \frac{h(i, c \lambda_n)}{E - (\epsilon_c + \hbar \omega_{\lambda,n})}$  
for $\epsilon_c > \epsilon_F $
and
$ R^{coll}_{(c,\lambda n)_{lj}}= \sum_i x^{lj}_i  \frac{h(i, c \lambda_n)}{E - (\epsilon_c - \hbar \omega_{\lambda,n})}$  
for $\epsilon_c < \epsilon_F $.
The square of the single particle component of the dressed state  is then given by 
\begin{equation}
a^2_{lj} = \frac{1}{1+ \sum_{c,\lambda,n} (R^{coll}_{(c,\lambda n)_{lj}})^2}
 \end{equation}  
while the admixture of the solution with the $c, \lambda n$ configuration is given by 
\begin{equation}
a^2_{(c, \lambda n)_{lj}}= \frac{(R^{coll}_{(c,\lambda n)_{lj}})^2}{1+ \sum_{c,\lambda,n} (R^{coll}_{(c,\lambda n)_{lj}})^2}
\end{equation}

The above solution provides the first term of the series of rainbow diagrams dressing the single-particle states, taking into account
the effect of many -phonon configurations.
Higher order  terms can be   generated  iterating the solution self-consistently [3].
We have not attempted to perform such a  calculation
in the present paper, but
employ  empirical renormalisation,
choosing  the single-particle intermediate states so as to 
reproduce the (low-lying) experimental energies, and eventually the outcome of the clothing process, i.e. 
$\tilde \epsilon_{lj}$ and $\tilde \phi_{lj} $. In other words, the 
solution of the Dyson equation will be acceptable only if  the single-particle component  of the solution is in agreement with the 
input used for the intermediate basis, a requirement  imposing a severe self-consistent condition to our diagonalization process.
In the calculations we have added a correction to the energy of the intermediate particle-phonon configurations, to correct  
for Pauli principle violation associated with many-phonon states  (anharmonicities, see Fig. 1(IV) of the main text),
implicitly considered in the empirical renormalization.
 Such corrections are particularly important in the case of the renormalization of the $d_{5/2}$ states, because of  the conspicuous coupling 
 to the $s_{1/2} \otimes 2^+$ configuration. 
 
 \begin{figure}[h!]
\includegraphics[width=0.45\textwidth]{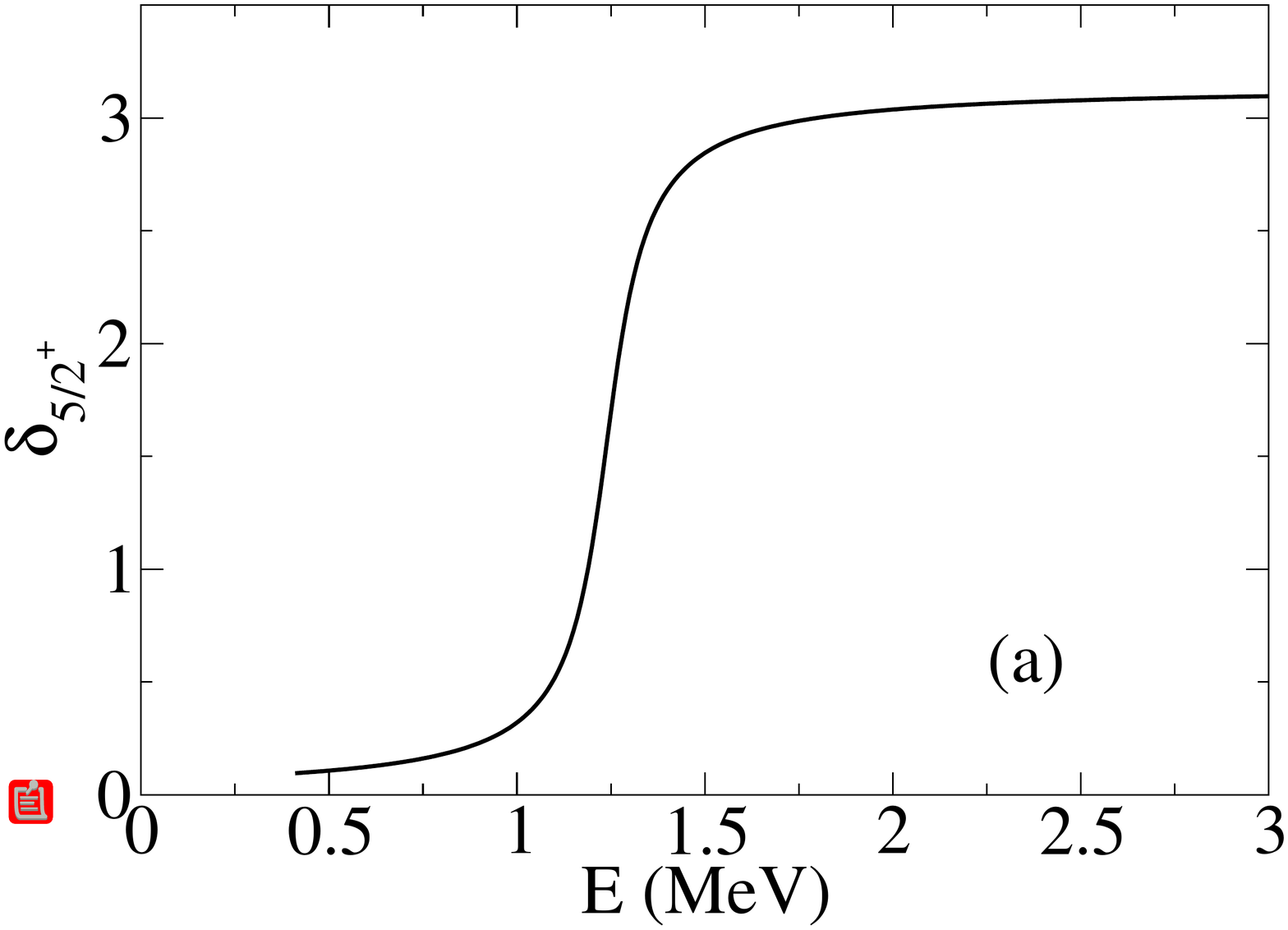}
\includegraphics[width=0.45\textwidth]{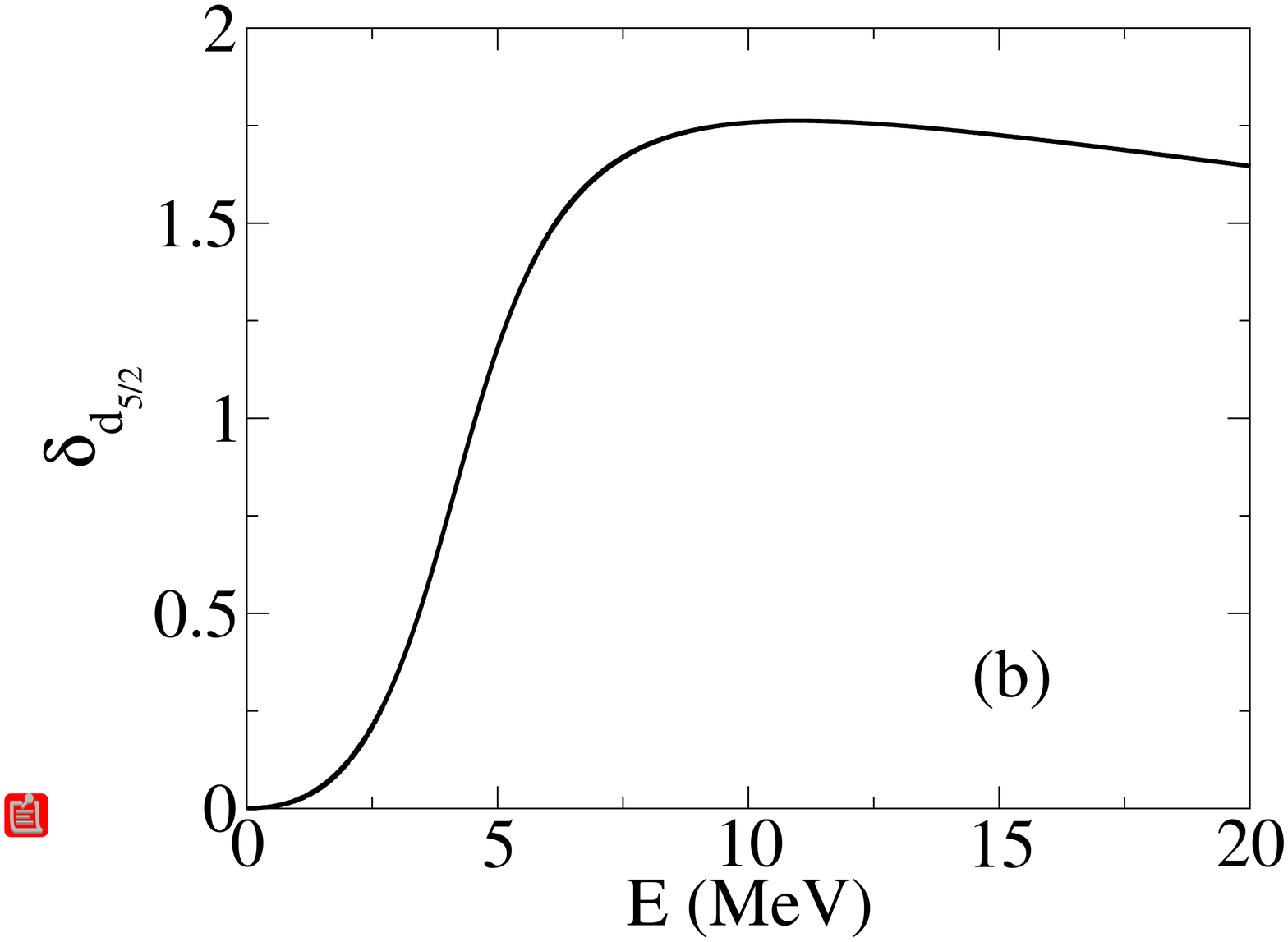}
\caption{(a) $5/2^+$ (NFT)$_{ren}$ phase shifts as a function of energy. 
(b) $d_{5/2}$ phase shifts   calculated with the bare potential.}
\label{fig:phase_shift}
\end{figure}

 The $2s_{1/2}$ orbital is not bound in our initial bare potential. As a consequence, if we use the
 same basis for the asymptotic  and the intermediate states, the coupling matrix element will be small due to the poor overlap in the corresponding 
 radial wavefunctions. Iterating the diagonalization, the $s_{1/2}$ state eventually  becomes bound, acquiring
 in the process  a collective component of the type $d_{5/2} \otimes 2^+$. At the same time,
 $5/2^+$  states  will acquire both  a bound collective component through the coupling with 
 the $s_{1/2} \otimes 2^+$ configuration and a localized (resonant) single-particle wave function due to the mixing of the various continuum states. 
 For economy, these processes are taken into account within the framework of  empirical renormalization 
 using a (intermediate) $s_{1/2}$ state bound by 0.5 MeV as in experiment, and verifying that this value coincides with the 
 energy $\tilde \epsilon_{1/2^+}$ of the  final dressed $| \widetilde{1/2^+}>$ state.

 The renormalised $5/2^+$ phase shift, displaying a resonance at $E_{res} =1.23 $ MeV of width $\Gamma= 160 $ keV,
 are shown in Fig. \ref{fig:phase_shift}, where  they are compared  wit the phase shifts calculated 
 with the  bare potential, displaying a broad resonance at $E= 6.5 $ MeV.

 \subsection{Bubble overcounting}

\begin{figure}[h!]
\includegraphics[width=0.5\textwidth]{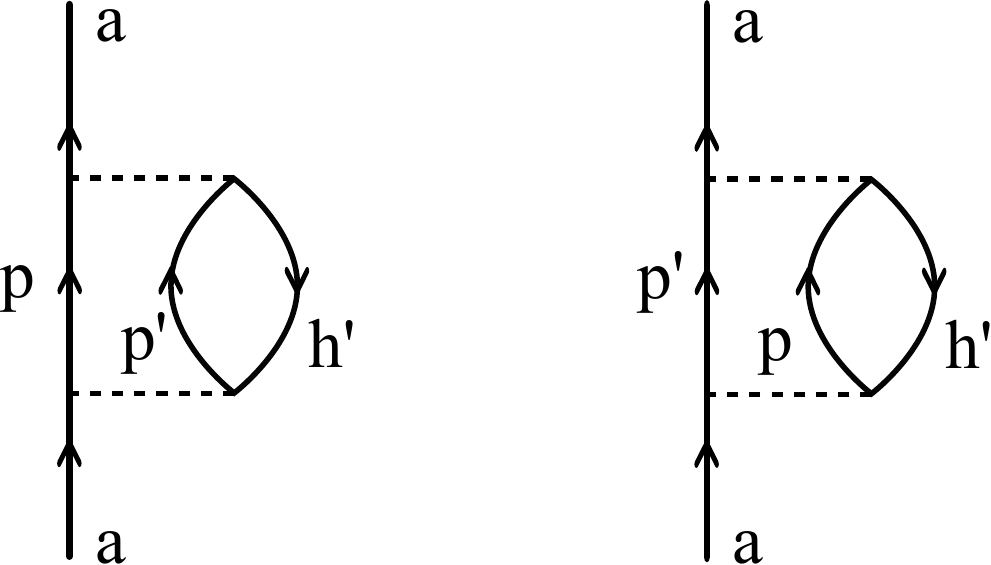}
\caption{The diagram at the right
 contains the same intermediate configuration as that of the left one.
If antisymmetrised two-body matrix elements are used, these 
two diagrams are identical.}
\label{fig:bubble_1}
\end{figure}

Freely summing over intermediate particle $ \otimes  $ phonon configurations when evaluating the 
self-energy diagrams,  leads as a rule   to some overcounting, since phonons 
are  linear combinations of particle-hole  configurations (see Fig.\ref{fig:bubble_1}) 

\begin{figure}[h!]
\includegraphics[width=0.45\columnwidth]{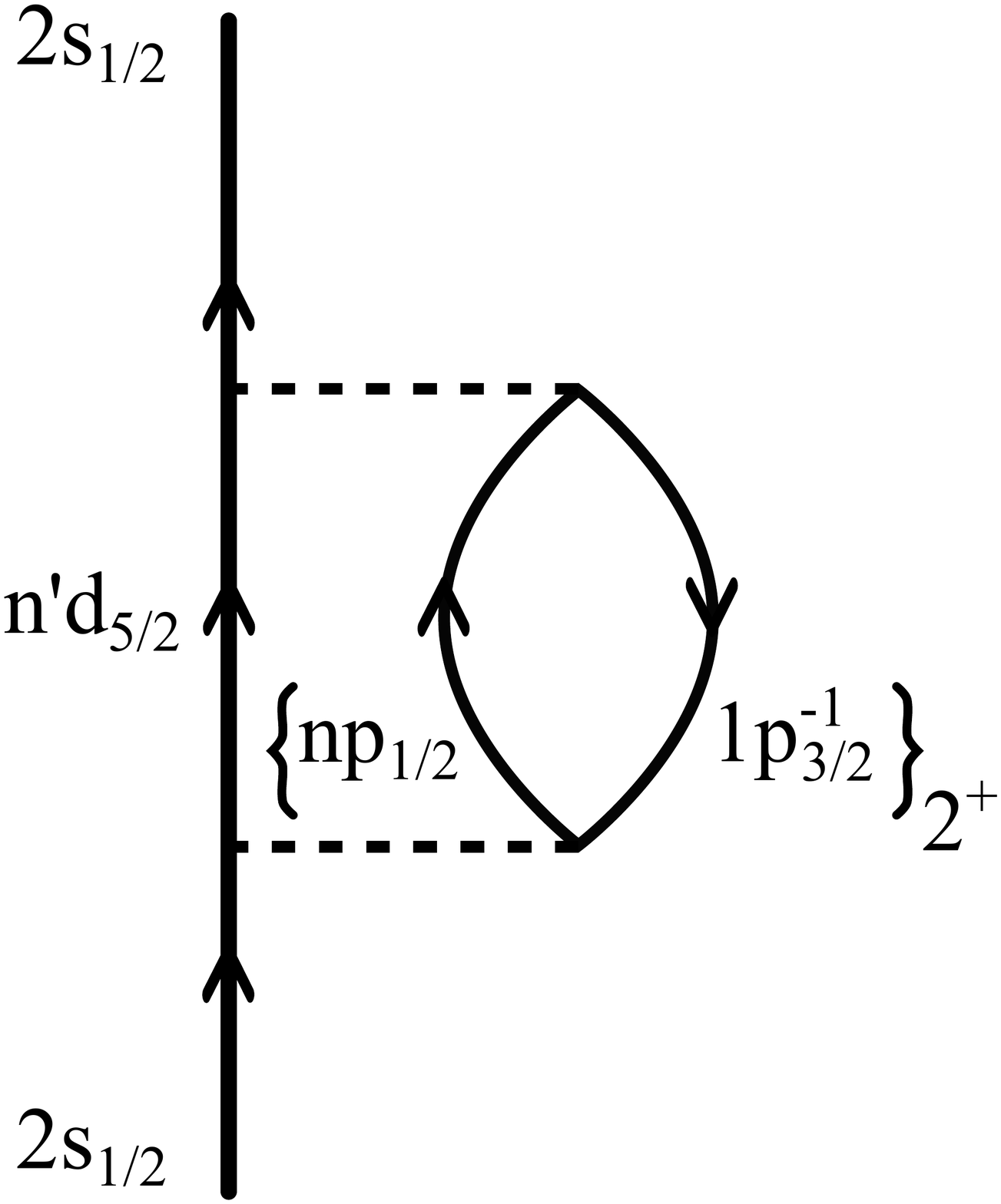}
\includegraphics[width=0.45\columnwidth]{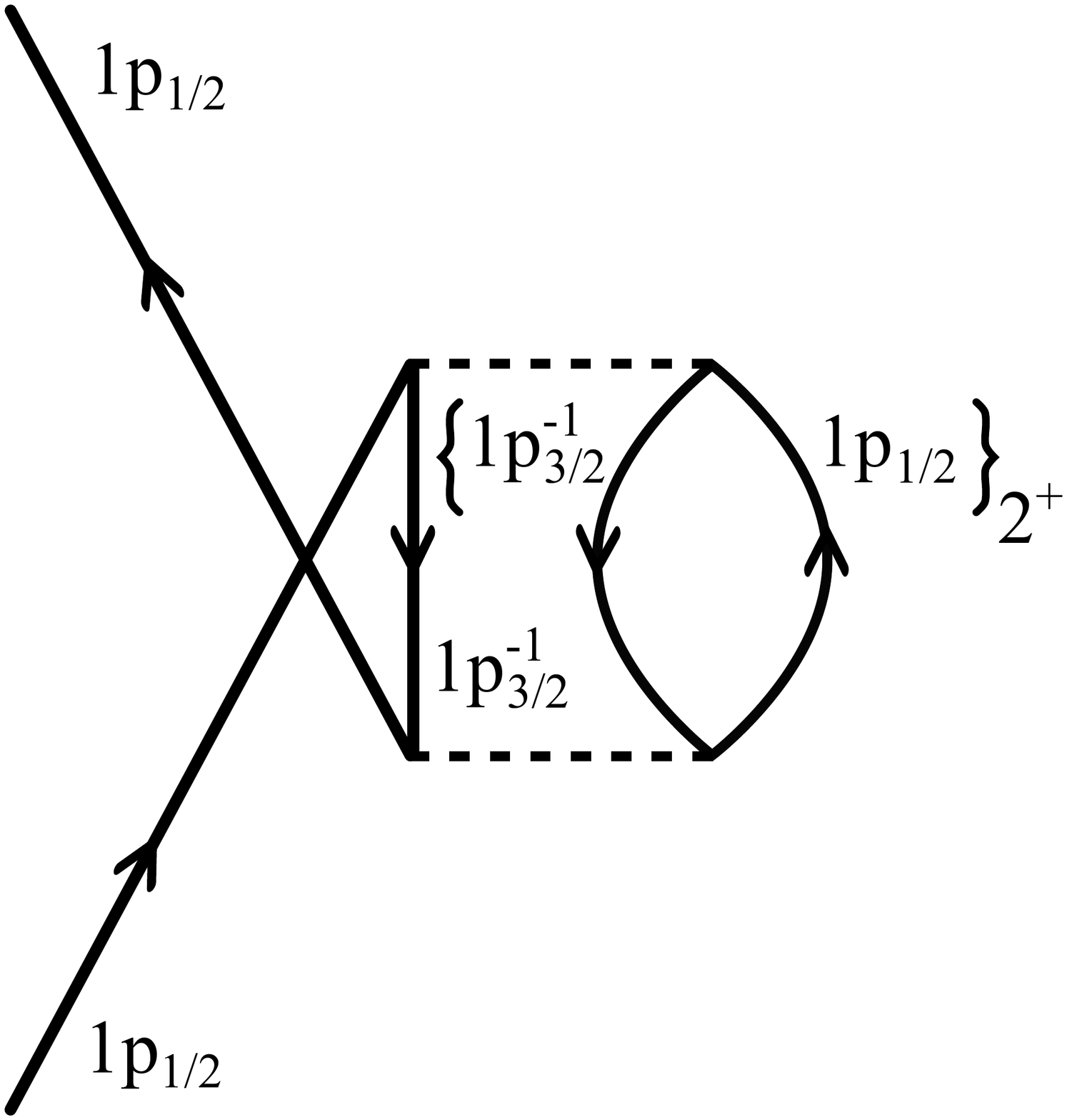}
\caption{ (left) Main p'-h'  configuration renormalizing the $2s_{1/2}$ state.
Summing over the number of nodes  $n$ and $n'$ does not 
lead to double counting, in keeping with the different parity and angular momentum
of the  $d_{5/2}$ and $p_{1/2}$ states. An identical reasoning applies to the $5/2^+$ states.
(right) Main p'-h'  configuration renormalizing the $2p_{1/2}$ state. 
 In this case only the  $1 p_{3/2}^{-1} $ hole state is taken into account,
and no overcounting is possible.}
\label{fig:bubble_2}
\end{figure}

Such overcounting can be cured 
 subtracting the overcounted bubble diagram from the 
particle-phonon (RPA series) calculation. 
In the present case, such overcounting is   small  
(a few per cent, smaller than our global theoretical  accuracy). In fact, the low lying $2^+$ phonon, 
which is by far  the most relevant phonon in our calculations, is constructed 
mostly out of  particle-hole transitions   between single-particle states of negative parity 
($1p_{3/2}^{-1} $ holes; $p_{1/2}$ (mainly), $p_{3/2}$ (very little), $f_{7/2}$ and $f_{5/2}$ (negligibly) particles) 
and marginally  by particle-hole transitions  of positive parity ($1s_{1/2}^{-1} $ holes, $d_{5/2}$ and $d_{3/2}$ particles). 
As a consequence, when evaluating the (positive parity) self-energies of the $s_{1/2}$ and $d_{5/2}$ states, 
the intermediate $p \otimes 2^+$ configuration  involves  particle states of positive parity, and  overcounting
is produced only by  the small positive parity particle-hole components of the $2^+$ phonon (see Figs. \ref{fig:bubble_2}).
On the other hand when evaluating the (negative parity) $p_{1/2}$ 
self-energy  no overcounting arises since only one $p_{3/2}$ hole contributes and no involved
summation can lead to overcounting (see Figs. \ref{fig:bubble_2}).

\subsection{Anharmonic effects}

 An important point concerns anharmonic effects in many-phonon configurations, associated with the Pauli principle, 
  which are not taken into account in the rainbow series. They are particularly relevant in the present case  due to the low angular moment of the 
  single-particle states
  involved in the calculation. The intermediate configurations $\tilde d_{5/2} \otimes 2^+$  and $\tilde s_{1/2} \otimes 2^+$ implicitly
  contain a 2-phonon
  component and this involves an anharmonic effect associated with  violation of the Pauli principle. In NFT the Pauli principle is restored  by the so called butterfly diagram
  (and associated ones, see [4],
  Eqs. (28-35)),  which takes into account    the fermion exchange between the  microscopic p-h structures of the involved phonons.
  We have not calculated in detail the diagram but we have estimated the associated correction by performing a RPA calculation blocking the relevant p-h transitions.
  This is an approximate  method which  coincides with the exact evaluation of the butterfly diagram in the two-level model. We obtain an increase of the energy 
  of the two-phonon configuration of 2.7 MeV and a reduction of the deformation $\beta_2$ parameter of about 30\%. The influence of these changes on the empirical-particle-phonon configurations energy depends on how much two-phonon weight they carry, which in turn is an output of our calculations. 
  A consistent calculation leads to an increase of 2.4 MeV for the $d_{5/2} \otimes 2^+$ configuration and of 1.2 MeV for the 
  $s_{1/2} \otimes 2^+$ one. The difference between these two corrections reflects the fact that the calculated 
 $2^+$ phonon admixture in the dressed $5/2^+$ state  (50\%) is about twice the value  obtained for the $1/2^+$ state (20\%).
The $\tilde d_{5/2} \otimes 2^+$  and $\tilde s_{1/2} \otimes 2^+$ configurations corrected for the Pauli principle are those 
entering diagrams (I)(a) and (III)(a) and (b) of Fig. 1 of the main text (see Fig. (IV) for the corresponding corrections).

 \subsection{Calculation of the dipole matrix element}

The measured value of the transition strength associated with 
the dipole transition  between the first excited state and the ground state of
$^{11}$Be    is $B(E1)=  0.102 \pm 0.02$   e$^2$ fm$^2$ = 0.32  W.U. [5]
(1 W.U. = $1/4 \pi (3/4)^2 (1.2 A^{1/3})^2$ = 0.3188 e$^2$fm$^2$). 
The transition strength is defined as 
\begin{equation}
B(E1; I_i \to I_f)= \frac{1}{2I_i +1} | <I_f|| i  M(E1)|| I_1>|^2
\end{equation}
The leading contribution for the $1/2^+ \to 1/2^-$ transition is  obtained 
calculating the transition between the single-particle part of the renormalised  wave functions,
weighted with the appropriate single-particle factor and multiplied by the recoil effective charge $e_{eff}= - e Z/A = -4e /11$:
\begin{equation}
B(E1; I_i \to I_f)=  \frac{e^2_{eff}}{2I_i +1}  a^2_{1/2^+} a^2_{1/2^-} |M^0|^2
\end{equation}
where 
\begin{equation}
M^0 \equiv <1/2^- || i M(E1) || 1/2^+ > =  \sqrt{ \frac{3}{2 \pi}} \frac{-1}{\sqrt{3}} I(E1)  = - \sqrt{\frac{1}{2\pi}} I(E1),
\end{equation}
with $I(E1)=  \int dr \tilde \psi_{1/2^+} (r) r  \tilde \psi_{1/2-}(r)$. The amplitudes are equal to    $a^2_{1/2^+} $ = 0.83   and $a^2_{1/2^-} $  = 0.81
(Figs. 1 of the main text , Eqs.(1) and (2), upper box)
Numerically we find $I(E1) \approx 5$ fm,  and then obtain the zero-order result
\begin{equation}
B(E1; 1/2^+ \to 1/2^-) \approx  \frac{16}{121} \frac{1}{2}  0.67 \frac{1}{2 \pi} 25 \quad {\rm e}^2  {\rm fm}^2  = 0.17   {\rm e}^2  {\rm fm}^2 .
\end{equation}

\begin{figure*}
\includegraphics[width=0.9\textwidth]{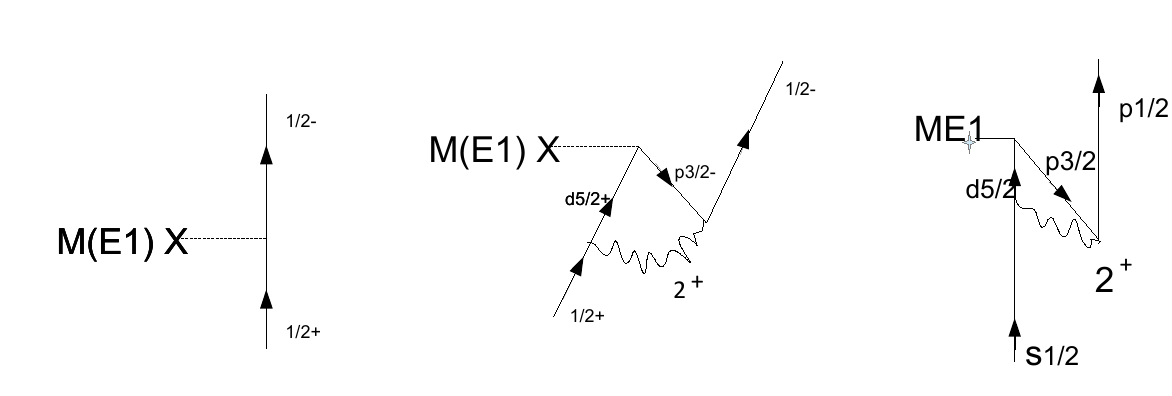}
\caption{Main processes contributing to the dipole transition between the first excited state and the ground state of
$^{11}Be$. The vertex correction associated with the low-lying $1^-$ strength (incipient GDPR) is estimated to be small.}
\label{fig:M1}
\end{figure*}

One has to consider, however, that one expects other important contributions from many-body processes.
The two time orderings associated with the most important ones
are shown in the right part of Figs. \ref{fig:M1}. They interfere in a destructive way with the  
leading contribution, leading to 
\begin{align}
B(E1; I_i \to I_f)=  \frac{e^2_{eff}}{2I_i +1}  a^2_{1/2^+} a^2_{1/2^-}  |  M^{0} +  M^{1(a)} + M^{1(b)} |^2= \nonumber \\
\frac{16}{121} \frac{1}{2}  0.66 \; | \sqrt{\frac{1}{2 \pi}}  \times  5  - 0.25  - 0.19 |^2  \quad {\rm e}^2  {\rm fm}^2
= 0.11 \quad {\rm e}^2  {\rm fm}^2. 
\end{align}
We remark  that the usual depletion of the low-lying E1 strength by the giant resonance is not very
effective in the present case, due to the poor overlap between the halo neutron single-particle states
and those of the nucleons of the core
(see [6],[7] 
p.2 and App. A,B; see also [8]
).
In fact, it can be estimated that the contribution of the giant dipole polarisation diagram is much smaller  than the bare 
M(E1) diagram due to the halo character of the  wave functions, and that it also much smaller than the  second order
quadrupole polarisation corrections  due
to the smaller deformation parameter of the GDR compared to quadrupole mode,  and to its  much larger excitation energy.

\subsection{Calculation of the charge radius}

An estimate of the 
difference between the value of the charge radius in $^{11}$Be and $^{10}$Be can be obtained 
by considering   that the dressed $ \widetilde{1/2^+}$ wave function has a $s_{1/2}$ single-particle part 
with amplitude  $a_{\widetilde{1/2^+}}= \sqrt{0.83}$ and a collective part dominated by the admixture with the lowest 
$2^+$ vibration of amplitude $a_{(d_{5/2^+} \otimes 2^+)_{1/2^+}} = \sqrt{0.17}$. 
The contribution to the square charge  radius due to the single-particle part is due to core recoil: 
\begin{equation}
(a_{\widetilde{1/2^+}})^2  \left(<r^2>_{^{10}Be} + \left( \frac{<r^2>_{s_{1/2}}^{1/2} }{11} \right)^2 \right) 
\end{equation} 
The contribution of the collective part is instead given by 
\begin{equation}
a^2_{(d_{5/2^+} \otimes 2^+)_{1/2^+}}   \left(<r^2>_{^{10}Be} (1+ \frac{2}{4 \pi} \beta^2) + \left( \frac{<r^2>_{d_{5/2},coll}^{1/2}} {11} \right)^2 \right),
\end{equation} 
where we have used the fact that  the radius of the  
nucleus  in its $2^+$ state (first excited state of a harmonic oscillator)
 is a factor ($1+ \frac{2}{4 \pi} \beta^2$) larger than in its ground state, while the neutron in the 
 $(d_{5/2^+} \otimes 2^+)_{1/2^+} $ state 
 is described by the wave function   $\psi_{d_{5/2},coll}$. 
One then obtains 
\begin{align} 
 <r^2>_{^{11}Be}  =  <r^2>_{^{10}Be} + 
(a_{1/2^+})^2    \left( \frac{<r^2>_{s_{1/2}}^{1/2}} {11} \right)^2 +  \nonumber   \\
 a^2_{(d_{5/2^+} \otimes 2^+)_{1/2^+}}
 \left(  <r^2>_{^{10}Be} \frac{2 \beta^2}{4 \pi} +    
       \left( \frac{<r^2>_{d_{5/2,coll}}^{1/2}} {11} \right)^2 
                                                                               \right ) .
\end{align} 
  Introducing the values $ <r^2>_{^{10}Be}  = $ 5.57 $fm^2$,    $(a_{1/2^+})^2  = 0.83, 
  a^2_{(d_{5/2^+} \otimes 2^+)_{1/2^+}} = 0.17,
   \beta_2 =1.2$, $<r^2>^{1/2}_{s_{1/2}} = 7.1 $ fm 
 and  $<r^2>^{1/2}_{d_{5/2},coll} = 3 $ fm   one obtains   $<r^2>_{^{11}Be} $ = 5.57 + 0.83 $\times$ 0.42
 + 0.17 $\times  $ (1.27 + 0.07) = 6.15 fm$^2$, and $ (<r^2>_{^{10}Be})^{1/2}$ = 2.48 fm,  
to be compared with the experimental value 2.46 fm $\pm $ 0.015 [9].

 \subsection{Coupled  equations }
 
An approximate solution of the Schr\"odinger equation $H \Psi_{a} = E \Psi_{a}$ for the 
 wave function of the clothed odd nucleon  can be expressed as 

\begin{equation}
\Psi_{a}=[\tilde \phi_{l_aj_a}, + [\tilde \phi^{coll}_{l_bj_b\lambda}, \cdot  \Gamma^{\dagger}_\lambda]_{j_a} 
- \bar{\tilde \phi}_{l_aj_a}, - [\bar{\tilde \phi}^D_{l_cj_c \lambda} \cdot  \Gamma_{\lambda}]_{j_a} ] \Phi^A_{GS}
\end{equation}
where $\Phi^A_{GS}$ denotes the ground state of the nucleus of  even mass number   $A$ 
(containing the correlations 
needed  so that it is the vacuum of the different elementary modes of excitation used as a basis to describe 
$^{11}$Be (i.e. $a_j | \Phi^A_{GS}> = \Gamma_{n\lambda} |\Phi^A_{GS}> =0)$,
 $\Gamma^{\dagger}_{\lambda} $ denotes a general  creation operator 
 of a vibrational state (phonon),  calculated using e.g. RPA,

\begin{equation}  
{\tilde \phi}_{l_aj_a}  = {R^x/r}_{l_a j_a} \Theta _{l_a j_a} 
\end{equation}

creates a particle in $l_a, j_a$,
\begin{equation}
 [\tilde \phi^{coll}_{l_bj_b\lambda} \cdot  \Gamma^{\dagger}_{\lambda}]_{j_a}  =
 (R^C_{b \lambda} (r) /r) [\Theta_{j_b} \cdot \Gamma^{\dagger}_{\lambda} ]_{j_a} 
 \end{equation}
creates a  particle-phonon state coupled to $j_a$ and parity $(-1)^{l_a} $,

\begin{equation}  
\bar{\tilde \phi}_{l_aj_a}  = (R^y_{l_a j_a}/r) \Theta _{l_a j_a} 
\end{equation}
annihilates a hole in $l_a, j_a$ and 

\begin{equation}  
[\bar{\tilde \phi}_{l_cj_c\lambda} \cdot  \Gamma_{\lambda}]_{j_a} = (R^D_{c \lambda}(r)/r) [\Theta _{j_c} \cdot \Gamma_{\lambda}]_{j_a}, 
\end{equation}
annihilates a hole-phonon coupled to $j_a$ and parity $(-1)^{l_a} $,
where the radial wavefunctions $R^x_a$ and $R^C_{b \lambda}$  must be  expanded  over a set of single-particle states
(HF) lying above  the Fermi energy:
\begin{equation}
R^x_a(r) = \sum_i x^{a_i} R^{HF}_{a_i}(r) \quad, \quad
R^C_{b \lambda}(r) = \sum_i R^{coll}_{b_i, \lambda}  R^{HF}_{b_i}(r); \epsilon_{a_i},\epsilon_{b_i} > \epsilon_F 
\label{pauli1}
\end{equation}
while 
$R^y_a$ and $R^D_{c \lambda}$ must be expanded over the occupied states:
\begin{equation}
R^y_a(r) = \sum_i y^{a_i} R^{HF}_{a_i}(r) \quad , \quad
R^D_{c \lambda}(r) = \sum_i   R^{coll}_{c_i, \lambda}  R^{HF}_{c_i}(r); \epsilon_{a_i},\epsilon_{c_i} < \epsilon_F .
\label{pauli2}
\end{equation}

The radial wave function  $R^D_{c \lambda}(r) $ 
 accounts for the proper antisymmetrization  of the   RPA ground state.  In fact the RPA ground state contains 2p-2h configurations,  
what implies that the odd particle will find  states inhibited by the Pauli principle  also above $\epsilon_F$. Reciprocally
the $R^y$ wave accounts for the possibilty that the impinging particle will find available states below $\epsilon_F$.

It can be  shown  that the radial wave functions satisfy the coupled equations
 \begin{eqnarray}
 [- \frac{\hbar^2}{2m}
\frac {\partial^2} {\partial r^2} + V_a(r )+ 0 \hbar   \omega]R^x_a(r) + 
\Xi_{a,b\lambda }(-rdV/dr)
 R^C_{b \lambda} (r)  \nonumber \\
 -\Xi_{a,c\lambda }(-rdV/dr)
 R^D_{c \lambda}(r) = E R^x_a(r) \nonumber
 \end{eqnarray}
\begin{eqnarray}
\Xi_{a,b\lambda }(-rdV/dr)
 R^x_a(r) + [- \frac{\hbar^2}{2m}
\frac {\partial^2} {\partial r^2}+ V_b(r)+ 1 \hbar   \omega]R^C_ {b \lambda}(r) \nonumber \\
-\Xi_{a,b\lambda }(-rdV/dr)
 R^y_a(r)
= E R^C_{b \lambda}(r) 
\nonumber 
\end{eqnarray}
\begin{eqnarray}
\Xi_{a,b\lambda }(-rdV/dr)
 R^C_{b \lambda} (r)
-[- \frac{\hbar^2}{2m}
\frac {\partial^2} {\partial r^2} + V_a(r )+ 0 \hbar   \omega]R^y_a(r) \nonumber \\
 +\Xi_{a,c\lambda }(-rdV/dr)
 R^D_{c \lambda}(r) = - E R^y_a(r) \nonumber
 \end{eqnarray}
 \begin{align}
\Xi_{a,c\lambda }(-rdV/dr) R^x_a(r) - 
[- \frac{\hbar^2}{2m}
\frac {\partial^2} {\partial r^2} + V_c(r)- 1 \hbar   \omega]R^D_{c \lambda}(r)  \nonumber \\
+ \Xi_{a,c\lambda }(-rdV/dr) R^y_a(r)
=  - E R^D_{c \lambda} (r) 
\end{align}
where 
\begin{equation}
V_a(r) = V(r) +V_{ls}(r) + V_{cent}(r)
\end{equation}
and
\begin{equation}
\Xi_{a,b\lambda } =  \langle \Theta _{j_a m_a} \sum_{\lambda \mu } \beta_{\lambda} Y_{\lambda \mu }
[\Gamma^{\dagger }_{\lambda \mu }+(-1)^{\mu} \Gamma_{\lambda \mu }] [\Theta _{j_b} \cdot \Phi_\lambda]_{j_a m_a}\rangle
\end{equation}

The conditions (\ref{pauli1})  and (\ref{pauli2}) imply that these equations must be solved using projection techniques.

 \subsection{Calculation of the absolute differential cross sections}

 \begin{figure}[b!]
\includegraphics[width=0.5\textwidth]{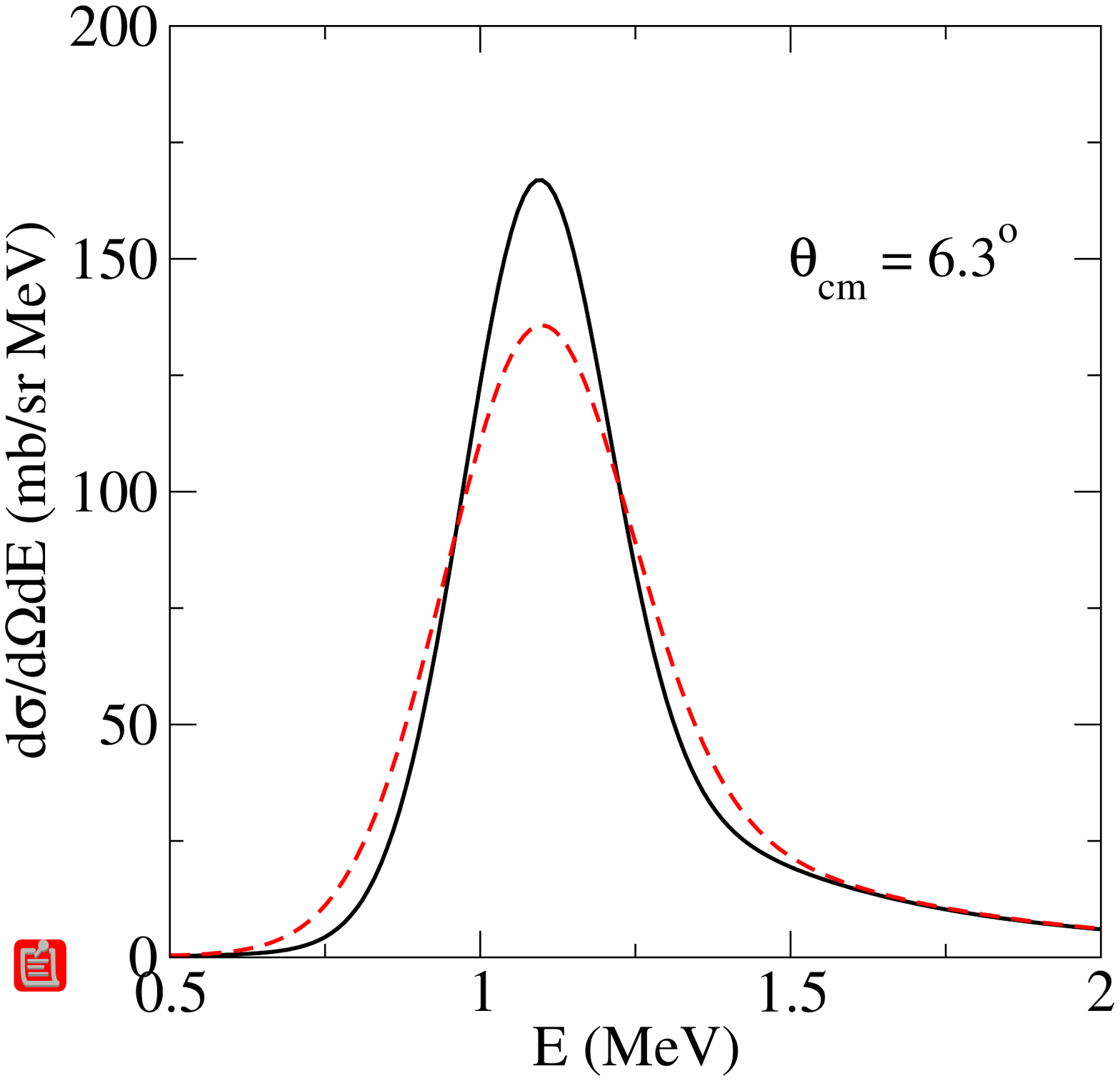}
\caption{The theoretical absolute  double differential cross section calculated for the reaction  
$^{10}$Be(d,p)$^{11}$Be populating the $5/2^+$ state at $\theta_{cm}$ = 6.3$^o$ is shown by the solid line.
The dashed line shows the result of convoluting this curve  with a Gaussian curve of FWHM $= 220$ keV
(estimated experimental resolution). 
}
\label{fig:width}
\end{figure}

The calculations of the angular distribution of the reaction  $^{10}$Be(d,p)$^{11}$Be($J_f^{\pi}$)  
have been performed in the post representation for the final  bound states  $J_f^{\pi}= 1/2^+$ and $1/2^-$ while
the prior representation has been adopted for the unbound state  $5/2^+$, to ensure a rapid convergence.
The transfer form factor is obtained from the single-particle  component of the many-body wave function.
In the  case of the $5/2^+$ final state, the $^{10}$Be-n potential  $V_{10Be-n}$ is also required. We have approximated $V_{10Be-n}$
with the bare Saxon-Woods potential of increased depth (bringing it to $V_{WS}$ = - 90 MeV), so that the  associated $d_{5/2}$ elastic
 phase shift displays a resonance for $E\approx  1.20 $ MeV. The angular distribution is obtained by summing the cross sections associated with 
each of  the $5/2^+$ eigenstates calculated  in a given box  and lying in the region of the resonance   (discretized continuum).
 Convergence  is obtained for boxes of the order of 40 fm. We note that the cross section associated with each 
 individual peak of energy $E_n$ approximately represents the value of the cross section integrated over an energy interval  lying between
 $(E_{n-1} + E_n)/2$ and $(E_n + E_{n+1})/2$.  In order  to compute  the energy distribution $d^2 \sigma /dE d\Omega$ for a   given value of $\theta$
 on  a sufficiently fine energy mesh,
 we have adopted  a continuum energy distribution with a Voigt shape, fitting the parameters   
  so that the integrals of the distribution  best reproduce the values  obtained in the different boxes in the appropriate intervals.
 We have folded the theoretical calculation with a gaussian curve of $FWHM=$ 220 keV, representing the experimental resolution,
 estimated from the width of the peaks of the discrete states (cf. Fig. \ref{fig:width}).
 We note that the experimental cross sections contain a background caused by the other partial waves, that we have not considered
 in our present calculation. 
 Finally, we have also calculated the angular distribution
associated with the one-nucleon transfer reaction $^{11}$Be(p,d)$^{10}$Be populating the
$2^+$ state in $^{10}$Be in the post representation, making use of the collective part  $R^C_{d_{5/2},2+}$  of the initial
 $1/2^+$ state  in $^{11}$Be (see Sect. 4).
 
 \vskip2cm

[1] A.Bohr and B.R.Mottelson, {\it Nuclear Structure} {\bf Vol. I}, Benjamin, New York (1969)

[2] H. Iwasaki{\em \ et al.},  
 {\it Phys.Lett.} {\bf B481}, 7 (2000)
 
[3] A. Idini, F. Barranco and E. Vigezzi, Phys, Rev. {\bf C 85} (2012) 014331

[4] D.R. Bes, R.A. Broglia, G.G. Dussel, R.J. Lotta and H.M. Sof\i a,
Nucl. Phys. A {\bf 260} (1976) 1

[5] E. Kwan  {\em \ et al.},   
Phys. Lett.  {\bf B732} (2014) 210

[6] F. Barranco,  P.F. Bortignon, R.A. Broglia, G. Col\`o and E. Vigezzi,  
Eur. Phys. J. A {\bf 11}, 385  (2001)

[7] R.A. Broglia, P.F. Bortignon, F. Barranco, E. Vigezzi, A. Idini and G. Potel, 
 Phys. Scr. {\bf 91} (2016) 063012
 
[8] I. Hamamoto and  S. Shimoura,   
J. Phys.  {\bf G34} (2007) 2715

[9] A. Krieger et al.,
Phys. Rev. Lett. {\bf 108} (2012) 142501


\end{document}